\documentclass[12pt]{article}
\def\Dj{\hbox{D\kern-.73em\raise.30ex\hbox{-}\raise-.30ex\hbox{}}}
\def\dj{\hbox{d\kern-.33em\raise.80ex\hbox{-}\raise-.80ex\hbox{\kern-.40em}}}

\usepackage{epsfig}
\usepackage{amsmath,amsthm,amsfonts,amssymb,amscd,cite}

\allowdisplaybreaks

\begin{document}

\baselineskip=0.30in

\vspace*{15mm}

\begin{center}
{\Large \bf On Relative Stabilities of Distinct Polyenes. An Extension of the
Concept of Conjugated Paths}

\vspace{10mm}

{\large \bf Viktorija Gineityte}

\vspace{9mm}

\baselineskip=0.20in
{\it Institute of Theoretical Physics and Astronomy,\\
Vilnius University,\\
Gostauto 12, LT-01108 Vilnius, Lithuania} \\
{\tt Viktorija.Gineityte@tfai.vu.lt}

\vspace{6mm}

\end{center}

\vspace{6mm}

\baselineskip=0.20in

\noindent {\bf Abstract }

\vspace{3mm}

{\small The study continues the previous development [MATCH, 72 (2014) 39-73] of
the perturbative approach to relative stabilities of pi-electron systems of
conjugated hydrocarbons modeled as sets of weakly-interacting initially-double
(C=C) bonds. Distinct isomers of acyclic hydrocarbons (polyenes) are now under
focus. The relevant total pi-electron energies ($\mathcal{E}$) are expressed in
the form of power series containing members ($\mathcal{E}_{(k)}$) of even orders
($k=0,2,4,...$) with respect to the averaged resonance parameter of
initially-single (C-C) bonds. Terms to within the sixth order ($k=6$) inclusive
are shown to be of importance for discrimination between similar isomers. In
this connection, missing expressions for corrections $\mathcal{E}_{(6)}$ are
originally derived. Conjugated paths of various lengths (i.e. linear chains
consisting of C=C and C-C bonds alternately) are shown to be the most important
(but not the only) fragments contributing to stabilization of any acyclic
pi-electron system. Again, new types of fragments (substructures) are revealed
(viz. the so-called composite conjugated paths) that contribute to
destabilization of the system concerned. As a result, formation of the total
energy of an acyclic pi-electron system is concluded to be governed by an
interplay between stabilizing and destabilizing factors. Accordingly, the
perturbative approach applied offers us an extension of the concept of
conjugated paths. Particular isomers containing four, five and six C=C bonds are
considered in a detail as examples.}

\vspace{5mm}

\baselineskip=0.30in

\section{Introduction}

Qualitative intuition-based concepts and models play a crucial role in chemistry
throughout its history. Accordingly, attempts to derive them deductively from
more sophisticated quantum-chemical approaches contribute to our understanding
of the nature of the given concept and/or model, as well as indicate directions
for its possible extensions and improvements.

The concepts of conjugated paths [1] and circuits [2-4] are successfully
applied to evaluate relative stabilities of pi-electron systems of similar
conjugated hydrocarbons [5], e.g. of various isomers of polyenes and of
different Kekul\'{e} valence structures of a certain benzenoid,
respectively. Chains and cycles consisting of double (C=C) and single (C-C)
bonds alternately are regarded here as the principal substructures
determining stabilities of the structures concerned. Some limitations of
these concepts also have been reported [5-8]. Difficulties in discriminating
between stabilities of isomers of extended polyenes characterized by
slightly different types of branching [5] are especially noteworthy in the
context of the present study.

In general, interpretation of stability of a certain pi-electron system
depends on the model applied. Given that the latter coincides with the
molecular graph [5, 9-11], the vertices and edges of which represent carbon
atoms and carbon-carbon bonds, respectively, the relevant total energy is
discussed in terms of properties of this graph (see e.g. [12,13]). Another
alternative consists in modeling of a conjugated system as a set of
weakly-interacting initially-double (C=C) bonds and thereby in employment of
the perturbation theory to evaluate and to rationalize relative stabilities
of isomers. Although the second option traditionally refers to acyclic
conjugated hydrocarbons [14], adequacy of perturbative approaches to
individual Kekul\'e valence structures of benzenoids also is beyond any
doubt [15]. Again, an evident analogy between the perturbational perspective
to pi-electron systems and that underlying the concepts of conjugated paths
and circuits gives us a hint about feasibility of a perturbation- theory-
based derivation of these concepts followed by their extensions (if
necessary).

The above-formulated task, however, is not easily achievable. Difficulties
in application to extended conjugated hydrocarbons of the most popular
perturbational molecular orbital (PMO) theory [14] may be mentioned among
the principal reasons here. Indeed, the original Dewar formula for the
second order energy ($\mathcal{E}_{(2)}$) underlying this theory yields
coinciding stabilities of numerous important pi-electron systems of distinct
constitutions, including different Kekul\'{e} valence structures of
benzenoids and isomers of polyenes. To circumvent this difficulty, reference
structures of artificial and involved nature are invoked instead of sets of
C=C bonds, e.g. two allyle radicals for a Kekul\'{e} valence structure of
benzene [14]. Such an option, however, makes the overall approach even more
remote from the concepts of conjugated paths and circuits.

An alternative to the standard PMO theory has been suggested recently [15,16],
wherein corrections of higher orders ($\mathcal{E}_{(3)},\mathcal{E}_{(4)},$
etc.) of the power series for total energies ($\mathcal{E}$) have been taken
into consideration along with second order ones ($\mathcal{E}_{(2)}$) instead of
passing to the above-discussed artificial reference structures. Thus, the
classical model of conjugated hydrocarbons as sets of weakly-interacting C=C
bonds has been preserved in this new approach. At the same time, the latter
exhibited a much higher discriminative potential as compared to that of the
standard PMO theory (which was shown to depend upon the number of corrections
actually included). Besides, compact and chemically illustrative expressions for
corrections $\mathcal{E}_{(3)},\mathcal{E}_{(4)}$ [15-17] and
$\mathcal{E}_{(5)}$ [15] have been derived using an original matrix form of the
perturbation theory (PT), namely the so-called non-commutative
Rayleigh-Schr\"odinger perturbation theory (NCRSPT) [18-20]. Application of the
latter approach to individual Kekul\'e valence structures of benzenoid
hydrocarbons has been carried out in the recent study [15]. Contributions both
of linear (acyclic) and of cyclic conjugated fragments (substructures) were
shown to be taken into consideration on the unified basis in the power series
for total energies. This implies the approach employed to embrace perturbative
analogues of both conjugated paths and circuits formally present in the
structures concerned. The above-mentioned contributions, however, are not easily
separable one from another. That is why no attempts were made to extend the
qualitative concepts either of conjugated paths or of circuits in this study.

The present article addresses a more simple case of acyclic conjugated
hydrocarbons (polyenes) containing no conjugated circuits. Our aim now consists
in a deductive revealing the principal types of conjugated fragments
(substructures) contributing to the relevant total pi-electron energies and
thereby in justifying and/or extending the concept of conjugated paths. At the
same time, the extension being sought might be relevant also to numerous related
concepts, such as conjugated paths invariants [21], the mean length of
conjugated paths [22], conjugation paths used in studies of donor/acceptor
functionalized pi-electron systems [23,24], etc. To achieve the above-formulated
end, the same perturbative approach will be applied along with the experience of
Ref.[15]. In respect of the latter, the following points deserve mentioning:
First, energy corrections of odd orders proved to take non-zero values for
conjugated- circuits- containing systems only. Thus, we may now confine
ourselves to members of the power series of even orders only. Second, energy
increments $\mathcal{E}_{(0)}$ and $\mathcal{E}_{(2)}$ were shown to be
determined by total numbers of C=C and C-C bonds of the given structure,
respectively. Thus, these corrections are expected to take coinciding values for
isomers of the same hydrocarbon. In this connection, corrections at least of
fourth and sixth orders seem to be required to discriminate between stabilities
of these isomers. Thus, formulae for sixth order energies are originally derived
in the Appendix of the present study.

The paper starts with a brief overview of the principal expressions of the
approach to be applied (Sect. 2). Thereupon, we turn to revealing the
principal types of conjugated fragments that contribute to the energy
corrections $\mathcal{E}_{(4)}$ and $\mathcal{E}_{(6)}$ (Sect. 3). The
final section is devoted to relative stabilities of specific polyenes.

\vspace{4mm}

\section{Expressions for energy corrections}

As already mentioned, acyclic conjugated hydrocarbons (polyenes) will be
considered as sets of weakly-interacting initially-double (C=C) bonds.
Moreover, the systems concerned belong to even alternant hydrocarbons
(AHs) [9-11, 25,26]. The above-mentioned two points will be taken into
consideration when constructing the relevant H\"uckel type Hamiltonian
matrix ($\mathbf{H}$).

Let the pi-electron system of a certain polyene to be initially represented by
an $2N-$dimensional basis set of $2p_{z}$ AOs of carbon atoms $\{\chi \}$, where
$N$ stands for the total number of C=C bonds. These AOs will be assumed to be
characterized by uniform Coulomb parameters ($\alpha $) as usual and the
equality $\alpha =0$ will be accepted. As with the standard H\"uckel model (see
e.g. [9,25]), resonance parameters between AOs of chemically bound pairs of
atoms only will be assumed to take non-zero values. Further, let the basis set
$\{\chi \}$ to be divided into two $N$-dimensional subsets $\{\chi ^{\ast }\}$
and $\{\chi^{\circ }\}$ so that pairs of orbitals belonging to any chemical bond
(C=C or C-C) find themselves in the different subsets. This implies the non-zero
resonance parameters to take place in the off-diagonal (inter-subset) blocks of
the Hamiltonian matrix ($\mathbf{H}$). Accordingly, zero submatrices stand in
the diagonal (intra-subset) positions of the matrix $\mathbf{H}$ as it is
peculiar to AHs in general [9, 25,26]. Finally, let us enumerate the basis
functions in such a way that orbitals belonging to the same C=C bond acquire the
coupled numbers $i$ and $N+i.$ As a result, resonance parameters of these strong
bonds take the diagonal positions in the intersubset blocks of the matrix
$\mathbf{H}$. Uniform values of these parameters ($\beta $) also is among
natural assumptions here. Let our (negative) energy unit to coincide with
$\beta$ in addition. The usual equality $\beta =1$ then immediately follows.
Similarly, the averaged resonance parameter of weak (C-C) bonds will be denoted
by $\gamma $ and supposed to be a first order term vs. the above-specified
energy unit.

In summary, Hamiltonian matrices of pi-electron systems of polyenes
($\mathbf{H}$) take a common form that may be represented as a sum of zero
($\mathbf{H}_{(0)}$) and first order matrices ($\mathbf{H}_{(1)}$) including
parameters of C=C and C-C bonds, respectively, viz.
\begin{equation}
\mathbf{H}=\mathbf{H}_{(0)}^{{}}+\mathbf{H}_{(1)}^{{}}=\left\vert 
\begin{array}{cc}
\mathbf{0} & \mathbf{I} \\ 
\mathbf{I} & \mathbf{0}
\end{array}
\right\vert +\left\vert 
\begin{array}{cc}
\mathbf{0} & \gamma \mathbf{B} \\ 
\gamma \mathbf{B}^{+} & \mathbf{0}
\end{array}
\right\vert ,
\end{equation}
where $\mathbf{I}$ here and below stands for the unit matrix and the
superscript + designates the transposed\ (Hermitian-conjugate) matrix. It
deserves adding here that unit off-diagonal elements of the submatrix $\mathbf{B}$\ ($B_{ij}=1,i\neq j)$ correspond to C-C bonds, otherwise these
take zero values. Meanwhile, the diagonal elements of the same submatrix ($B_{ii}$) vanish because entire resonance parameters of C=C bonds are
included into the zero order matrix $\mathbf{H}_{(0)}$.

The Hamiltonian matrix of Eq.(1) coincides with that representing Kekul\'e
valence structures of benzenoids [15] because of similar constitutions of
both systems. Accordingly, the subsequent steps towards derivation of the
relevant energy corrections also are similar. Thus, we start with passing to
a new basis \{$\varphi $\} consisting of bonding and antibonding orbitals
of C=C bonds defined as normalized sums and differences of pairs of AOs $\chi _{i}^{\ast }$ and $\chi _{N+i}^{\circ }$\ and referred to below as bond
orbitals. The transformed Hamiltonian matrix then meets the requirements of
the NCRSPT (see the Appendix). As a result, general formulae for members of
the power series for total energies ($\mathcal{E}$) may be applied that have
been derived earlier [15-17] using this PT [18-20]. As already mentioned
(Sect.1), we confine ourselves to terms $\mathcal{E}_{(k)}$ of even orders ($k=0,2,$ $4,..$). Let us turn now to individual members of the power series.

The zero order energy ($\mathcal{E}_{(0)}$) coincides with $2N,$ whatever
the specific structure of the given system. The subsequent second order
member ($\mathcal{E}_{(2)}$) takes a rather simple form, viz.
\begin{equation}
\mathcal{E}_{(2)}=4Tr(\mathbf{G}_{(1)}\mathbf{G}_{(1)}^{+})>0,
\end{equation}
whereas the fourth order one ($\mathcal{E}_{(4)}$) consists of a sum of two
components [16]:
\begin{equation}
\mathcal{E}_{(4)}^{(+)}=4Tr(\mathbf{G}_{(2)}\mathbf{G}_{(2)}^{+})>0,\quad 
\mathcal{E}_{(4)}^{(-)}=-4Tr(\mathbf{G}_{(1)}\mathbf{G}_{(1)}^{+}
\mathbf{G}_{(1)}\mathbf{G}_{(1)}^{+})<0.
\end{equation}
The notation $Tr$ here and below stands for a \textit{Trace} of the whole
matrix product within parentheses, and $\mathbf{G}_{(1)}$ and $\mathbf{G}_{(2)}$ are the principal matrices of the NCRSPT [17-20] of the first and
second orders, respectively, specified below. As is seen from Eqs. (2) and
(3), \textit{Traces} of positive-definite matrices [27] of the type $\mathbf{AA}^{+}$ stand in these relations. Thus, sums of squares of elements of
matrices $\mathbf{G}_{(1)},$ $\mathbf{G}_{(2)}$ and $\mathbf{G}_{(1)}\mathbf{G}_{(1)}^{+}$\ are contained there. This implies an \textit{a priori}
positive sign of the second order energy $\mathcal{E}_{(2)}.$ Meanwhile,
the components of the fourth order correction $\mathcal{E}_{(4)}$\ are of
opposite signs as indicated by additional superscripts $(+)$ and $(-)$.

Let us now dwell on matrices $\mathbf{G}_{(1)}$ and $\mathbf{G}_{(2)}$
[17-20]. In the particular case of the NCRSPT employed in the present study
(see the Appendix), these matrices are expressible as follows 
\begin{equation}
\mathbf{G}_{(1)}=-\frac{1}{2}\mathbf{R,\quad G}_{(2)}=
-\frac{1}{2}(\mathbf{SG}_{(1)}-\mathbf{G}_{(1)}\mathbf{Q})
=\frac{1}{4}(\mathbf{SR}-\mathbf{RQ})=\frac{1}{4}(\mathbf{SR}+\mathbf{RS}),
\end{equation}
where matrices $\mathbf{S,Q}$ and $\mathbf{R}$ contain resonance parameters
between the above-specified bond orbitals (BOs) $\{\varphi \}$. Let bonding
BOs (BBOs) and the antibonding ones (ABOs) to be correspondingly denoted by
subscripts $(+)$ and $(-)$, e.g. $\varphi _{(+)i}$\ and $\varphi_{(-)l}$
will stand for the BBO of the Ith C=C bond and for the ABO of the Lth one,
respectively. Individual elements of matrices $\mathbf{S,Q}$ and $\mathbf{R}$
may be then explicitly expressed as follows
\begin{equation}
S_{ij}=<\varphi _{(+)i}\mid \widehat{H}\mid \varphi _{(+)j}>,Q_{lm}=<\varphi
_{(-)l}\mid \widehat{H}\mid \varphi _{(-)m}>,R_{il}=<\varphi _{(+)i}\mid 
\widehat{H}\mid \varphi _{(-)l}>,
\end{equation}
where the BOs concerned are shown inside the bra- and ket-vectors. At the
same time, the new matrices $\mathbf{S,Q}$ and $\mathbf{R}$ are related to
the principal submatrices ($\gamma \mathbf{B}$ and $\gamma \mathbf{B}^{+}$)
of our initial Hamiltonian of Eq.(1), viz. 
\begin{equation}
\mathbf{S}=-\mathbf{Q}=\frac{\gamma }{2}(\mathbf{B}+\mathbf{B}^{+}),\qquad 
\mathbf{R}=\frac{\gamma }{2}(\mathbf{B}^{+}-\mathbf{B}).
\end{equation}
It is seen that matrices $\mathbf{S(Q)}$ and $\mathbf{R}$\ are proportional
to the symmetric (Hermitian) and skew-symmetric (skew-Hermitian) parts of
the matrix \ $\mathbf{B}$, respectively. On this basis, $\mathbf{G}_{(1)}$
and $\mathbf{G}_{(2)}$ of Eq.(4) may be easily shown to be skew-symmetric
(skew-Hermitian) matrices [28]. After an additional invoking the
above-mentioned equality $B_{ii}=0$ for any $i$, we then obtain that 
\begin{equation}
S_{ii}=Q_{ii}=R_{ii}=G_{(1)ii}=G_{(2)ii}=0,
\end{equation}
i.e. matrices embraced by Eq.(7) contain zero diagonal elements. A formal
coincidence between matrices $\mathbf{S}$ and $-\mathbf{Q}$ also is seen
from Eq.(6). Just this circumstance allows us to eliminate the matric $\mathbf{Q}$ as shown in the last relation of Eq.(4). It deserves adding
finally that the matrix product $\mathbf{G}_{(1)}\mathbf{G}_{(1)}^{+}$
determining the negative component of the fourth order energy ($\mathcal{E}_{(4)}^{(-)}$) is a symmetric (Hermitian) matrix. Consequently, diagonal
elements $(\mathbf{G}_{(1)}\mathbf{G}_{(1)}^{+})_{ii}$\ take non-zero values
and prove to be responsible for a large part of this energy component [16].

The energy correction of the sixth order ($\mathcal{E}_{(6)}$) is derived in
the Appendix. Four components reveal themselves in this correction, viz.
\begin{equation}
\mathcal{E}_{(6)1}^{(+)}=4Tr(\mathbf{G}_{(3)}^{o}\mathbf{G}_{(3)}^{o+})>0,
\end{equation}
\begin{equation}
\mathcal{E}_{(6)2}^{(+)}=8Tr(\mathbf{G}_{(1)}\mathbf{G}_{(1)}^{+}\mathbf{G}
_{(1)}\mathbf{G}_{(1)}^{+}\mathbf{G}_{(1)}\mathbf{G}_{(1)}^{+})>0,
\end{equation}
\begin{equation}
\mathcal{E}_{(6)}^{(-)}=-32Tr[(\mathbf{G}_{(1)}\mathbf{G}_{(2)}^{+})(\mathbf{G}_{(1)}\mathbf{G}_{(2)}^{+})^{+}]\equiv -32Tr[(\mathbf{G}_{(1)}^{+}\mathbf{G}_{(2)})(\mathbf{G}_{(1)}^{+}\mathbf{G}_{(2)})^{+}]<0,
\end{equation}
\begin{equation}
\mathcal{E}_{(6)}^{(u)}=8Tr(\mathbf{G}_{(1)}\mathbf{G}_{(2)}^{+}\mathbf{G}
_{(1)}\mathbf{G}_{(2)}^{+})\equiv 8Tr[(\mathbf{G}_{(1)}\mathbf{G}_{(2)}^{+})(\mathbf{G}_{(2)}\mathbf{G}_{(1)}^{+})^{+}],
\end{equation}
where 
\begin{equation}
\mathbf{G}_{(3)}^{o}=-\frac{1}{2}(\mathbf{SG}_{(2)}-\mathbf{G}_{(2)}\mathbf{Q})=-\frac{1}{8}[(\mathbf{S)}^{2}\mathbf{R}+2\mathbf{SRS}+\mathbf{R(S)}^{2}].
\end{equation}
The superscript $o$ is used here to distinguish the above-introduced matrix 
$\mathbf{G}_{(3)}^{o}$ from the standard third order matrix of the NCRSPT $\mathbf{G}_{(3)}^{{}}$ defined by Eqs. (A5) and (A6).

The components of Eqs.(8) and (9) resemble $\mathcal{E}_{(4)}^{(+)}$ of
Eq.(3) in respect of both an \textit{a priori} positive sign and
skew-symmetric (skew-Hermitian) nature of underlying matrices $\mathbf{G}_{(3)}^{o}$ and $\mathbf{G}_{(1)}\mathbf{G}_{(1)}^{+}\mathbf{G}_{(1)}$,
respectively. These components are correspondingly designated by additional
subscripts $1$ and $2$. Accordingly, the only \textit{a priori} negative
component is shown in Eq.(10). The latter is alternatively expressible in
terms of matrix products either $\mathbf{G}_{(1)}\mathbf{G}_{(2)}^{+}$ or 
$\mathbf{G}_{(1)}^{+}\mathbf{G}_{(2)}$. Meanwhile, the sign of the last
component of the sixth order energy of Eq.(11) cannot be established \textit{a priori} and the superscript $(u)$ (undefined) is used.

Let us dwell now on interpretation of elements of the principal matrices
determining our energy increments of Eqs.(2) and (3) and (8)-(11). Let us
start with the simplest matrices $\mathbf{G}_{(1)},$ $\mathbf{G}_{(2)}$ and $%
\mathbf{G}_{(3)}^{o}.$\ As is seen from Eqs.(4) and (5), the element $%
G_{(1)il}$ connects the BBO $\varphi _{(+)i}$ and the ABO $\varphi _{(-)l}$.
Moreover, it is proportional to the relevant resonance parameter ($R_{il}$)
and inversely proportional to the energy gap between BBOs and ABOs (equal to 
$2$). Consequently, this element represents the direct (through-space)
interaction between BOs $\varphi _{(+)i}$ and $\varphi _{(-)l}$. Besides,
direct intrabond interactions $G_{(1)ii}$\ vanish (see Eq.(7)). Again, the
one-to-one correspondence between non-zero elements of the matrix $\mathbf{B}
$ and C-C bonds along with Eq.(6) allows us to expect non-zero direct
interactions ($G_{(1)il}\neq 0$) to refer to BOs ($\varphi _{(+)i}$ and $%
\varphi _{(-)l}$) belonging to first-neighbouring C=C bonds only, the latter
coinciding with those connected by a C-C bond. Further, the second order
elements $G_{(2)il}$ are accordingly interpretable as indirect
(through-bond) interactions of the same BOs. Indeed, from Eq.(4) we obtain%
\begin{equation}
G_{(2)il}=\frac{1}{4}[\mathop{\displaystyle \sum }\limits_{(+)j}S_{ij}R_{jl}-%
\mathop{\displaystyle \sum }\limits_{(-)m}R_{im}Q_{ml}],
\end{equation}
where sums over $(+)j$ and over $(-)m$ correspondingly embrace all BBOs and
all ABOs of the given system. It is seen that both BBOs ($\varphi _{(+)j}$)
and ABOs ($\varphi _{(-)m}$) of other bonds play the role of mediators here
[Note that $j\neq i$ and $m\neq l$\ because of Eq.(7)]. Moreover, the
orbitals $\varphi _{(+)j}$ and $\varphi _{(-)m}$ should overlap directly
both with $\varphi _{(+)i}$ and with $\varphi _{(-)l}$ to be efficient
mediators. That is why non-zero indirect interactions correspond to pairs of
second-neighbouring C=C bonds possessing a common first neighbour.
Analogously, the third order elements $G_{(3)il}^{o}$ represent the indirect
interactions of the same BOs by means of two mediators. Pairs of BOs ($%
\varphi _{(+)i},\varphi _{1}$), ($\varphi _{1},\varphi _{2}$) and ($\varphi
_{2},\varphi _{(-)l}$) should overlap directly in this case, where $\varphi
_{1}$ and $\varphi _{2}$\ stand for mediating orbitals.

Elements of matrix products determining the energy components $\mathcal{E}%
_{(4)}^{(-)},\mathcal{E}_{(6)2}^{(+)},$ $\mathcal{E}_{(6)}^{(-)}$ and $%
\mathcal{E}_{(6)}^{(u)}$\ also may be interpreted as indirect interactions
of BOs. For example, the element $(\mathbf{G}_{(1)}\mathbf{G}_{(1)}^{+}%
\mathbf{G}_{(1)})_{il}$\ represents a certain specific indirect interaction
between BOs $\varphi _{(+)i}$ and $\varphi _{(-)l},$ wherein the mediating
orbitals necessarily coincide with an ABO $\varphi _{(-)1}$ and a BBO $%
\varphi _{(+)2},$\ respectively, whilst the interaction itself consists of
three successive direct interactions $G_{(1)i1},G_{(1)12}^{+}$ and $%
G_{(1)2l} $ referring to pairs of BOs ($\varphi _{(+)i},\varphi _{(-)1}$), ($%
\varphi _{(-)1},\varphi _{(+)2}$) and ($\varphi _{(+)2},\varphi _{(-)l}$).
Non-zero values of these direct components evidently are required to ensure
a non-vanishing third order element $(\mathbf{G}_{(1)}\mathbf{G}_{(1)}^{+}%
\mathbf{G}_{(1)})_{il}.$\ Similarly, an element ($\mathbf{G}_{(1)}\mathbf{G}%
_{(2)}^{+})_{ij}$ involves a direct and an indirect interaction. Besides,
pairs of bonding BOs play the role of interacting orbitals for elements both
($\mathbf{G}_{(1)}\mathbf{G}_{(1)}^{+})_{ij}$ and ($\mathbf{G}_{(1)}\mathbf{G%
}_{(2)}^{+})_{ij},$ e.g. the element $(\mathbf{G}_{(1)}\mathbf{G}%
_{(1)}^{+})_{ij}$\ represents the indirect interaction between BBOs $\varphi
_{(+)i}$ and $\varphi _{(+)j}$\ via ABOs of the first-neighbouring C=C
bonds. It is evident that the mediating ABO should overlap with both $%
\varphi _{(+)i}$ and $\varphi _{(+)j}$ in this case too. Accordingly, the
diagonal element $(\mathbf{G}_{(1)}\mathbf{G}_{(1)}^{+})_{ii}$\ may be
interpreted as the indirect self-interaction of the BBO $\varphi _{(+)i}.$

In summary, the above analysis yields the following rule: Any matrix element
of the kth order connecting two BOs $\varphi _{s}$ and $\varphi _{t}$\ and
determining an energy component takes a non-zero value, if there is at least
a single non-zero product of resonance parameters, i.e. 
\begin{equation}
<\varphi _{s}\mid \widehat{H}\mid \varphi _{1}><\varphi _{1}\mid \widehat{H}%
\mid \varphi _{2}>...<\varphi _{k-2}\mid \widehat{H}\mid \varphi
_{k-1}><\varphi _{k-1}\mid \widehat{H}\mid \varphi _{t}>\neq 0,
\end{equation}%
where $\varphi _{1},\varphi _{2},...\varphi _{k-1}$\ stand for mediating
orbitals. Given that the condition of Eq.(14) is met, we will say that in
the given system there is a pathway of the (k-1)th order between BOs $%
\varphi _{s}$ and $\varphi _{t}$. In the case of diagonal elements, we will
accordingly have to deal with self-returning pathways. Besides, steps inside
the same C=C bond are not allowed in these pathways because of Eq.(7). It
also deserves emphasizing that the term \textit{a pathway }(over BOs) is
used here and below to make a distinction from conjugated \textit{paths}
defined in terms of chemical bonds.

After returning to the power series for total energies of Eqs.(2),(3) and
(8)-(11), we may finally conclude that the higher is the order parameter ($k$%
), the more extended fragment of the whole system generally is embraced by
the given correction ($\mathcal{E}_{(k)}$). In this respect, the present
series resembles the graph-theoretic cluster expansion for total energy
[29], as well as the expansion in terms of moments [30].

\vspace{4mm}

\section{Conjugated fragments contributing to total energies of polyenes}

As discussed already (Sect. 2) separate increments to total energies are
determined by matrices $\mathbf{G}_{(1)},\mathbf{G}_{(2)},$ $\mathbf{G}%
_{(3)}^{o},$ \ $\mathbf{G}_{(1)}\mathbf{G}_{(1)}^{+},$ $etc.$ Thus, we will
look for relations between elements of these matrices, on the one hand, and
conjugated fragments present in \ the given system, on the other hand. The
above-enumerated matrices are collected below into three groups that are
analyzed separately.

\subsection{Relations between elements of matrices $G_{(1)}$ and $G_{(2)}$
and the simplest conjugated paths}

Let us start with elements of the first order matrix $\mathbf{G}_{(1)}$\
defined by Eq.(4). An element $G_{(1)il}$ (as well as $G_{(1)li}$) takes a
non-zero value, if the Ith C=C bond and the Lth one (the underlying orbitals 
$\varphi _{(+)i}$ and $\varphi _{(-)l}$ belong to)\ are connected by a C-C
bond (Sect. 2). This implies non-zero elements $G_{(1)il}(G_{(1)li})$ to
correspond to butadiene-like fragments and thereby to individual simplest
conjugated paths (CPs) embracing two neighbouring C=C bonds and abbreviated
below as CP(2)s. Moreover, the above-specified significant elements are
local in their nature and, consequently, take uniform values for all CP(2)s.
Let us also recall that the matrix $\mathbf{G}_{(1)}$ gives birth to the
positive second order energy of Eq.(2). This implies all CP(2)s of the given
polyene to contribute uniform stabilizing increments to the energy $\mathcal{E}_{(2)}$, the latter then being proportional to the total number of these
paths.

\begin{figure}
\includegraphics[width=0.9\textwidth]{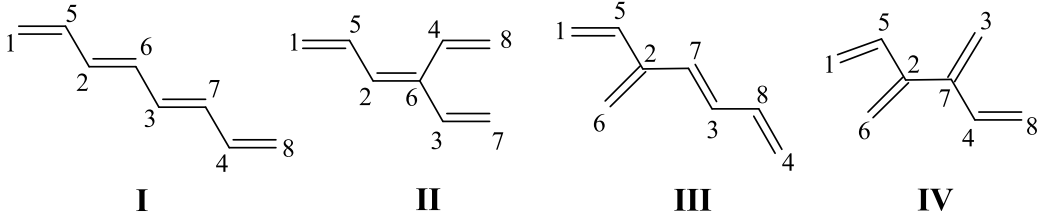}
\caption{Isomers of octatetraene (I-IV) containing four C=C bonds (N=4).
Numberings of $2p_{z}$ AOs of carbon atoms also are shown, where AOs under
numbers 1,2,3,4 and 5,6,7,8 belong to subsets $\{\chi^{\ast}\}$ and
$\{\chi^{\circ}\}$, respectively. Any initially-double bond C$_{K}$=C$_{N+K}
$ is supposed to acquire the number K, where K=1,2,3,4.}
\end{figure}

To exemplify the above simple rule, let us consider a linear polyene
containing N C=C bonds and its cross-conjugated counterpart (dendralene)
[31]. Carbon atoms and thereby the relevant 2p$_{z}$ AOs of these extended
systems are assumed to be enumerated as follows
\begin{gather*}
\mathrm{C}_{1}=\mathrm{C}_{N+1}-\mathrm{C}_{2}=\mathrm{C}_{N+2}
-\mathrm{C}_{3}=\mathrm{C}_{N+3}...-\mathrm{C}_{N}=\mathrm{C}_{2N},\\
\mathrm{C}_{1}=\mathrm{C}_{N+1}-\mathrm{C}_{2}(=\mathrm{C}_{N+2})
-\mathrm{C}_{N+3}(=\mathrm{C}_{3})-\mathrm{C}_{4}(=\mathrm{C}_{N+4})-...,
\end{gather*}
where the C=C bonds placed outside the principal chain of carbon atoms of
dendralene are shown within parentheses. Numbers 1,2...N and N+1, N+2...2N
refer here to subsets $\{\chi ^{\ast }\}$ and $\{\chi ^{\circ }\},$
respectively [see also Fig.1, where isomers of octatetraene I and IV serve
as examples of the systems concerned for N=4].

The principal first order matrices (viz. $\mathbf{B}$\ and $\mathbf{G}_{(1)}$)
of linear polyenes were shown to take a common form valid for any N [15]
[these unified representations have be\ en denoted by $\mathbf{B(}N\mathbf{)}$
and $\mathbf{G}_{(1)}\mathbf{(}N\mathbf{)}$]. After invoking
Eqs.(1),(4) and (6), analogous matrices are easily constructable also for
dendralenes. Let the latter to acquire additional superscripts $\prime $.
For comparison, matrices $\mathbf{G}_{(1)}\mathbf{(}N\mathbf{)}$\ and $%
\mathbf{G}_{(1)}^{\prime }\mathbf{(}N\mathbf{)}$\ are as follows%
\begin{equation}
\mathbf{G}_{(1)}\mathbf{(}N\mathbf{)}\mathbf{=}\mathbf{-}\frac{\gamma }{4}%
\left\vert 
\begin{array}{ccccc}
0 & 1 & 0 & 0 & ... \\ 
-1 & 0 & 1 & 0 & ... \\ 
0 & -1 & 0 & 1 & ... \\ 
0 & 0 & -1 & 0 & ... \\ 
.. & .. & .. & .. & ...%
\end{array}%
\right\vert ,\mathbf{G}_{(1)}^{\prime }\mathbf{(}N\mathbf{)}\mathbf{=}%
\mathbf{-}\frac{\gamma }{4}\left\vert 
\begin{array}{ccccc}
0 & 1 & 0 & 0 & ... \\ 
-1 & 0 & -1 & 0 & ... \\ 
0 & 1 & 0 & 1 & ... \\ 
0 & 0 & -1 & 0 & ... \\ 
.. & .. & .. & .. & ...%
\end{array}%
\right\vert ,
\end{equation}%
where a standard factor ($-\gamma /4$) is introduced in front of matrices
concerned for convenience. It is seen that two non-zero elements ($\mathbf{G}%
_{(1)il}$ and $\mathbf{G}_{(1)li}$) correspond to any C-C bond in these
matrices\ and thereby to any CP(2) of our polyenes, and these elements take
uniform absolute values in addition. Moreover, matrices $\mathbf{G}_{(1)}%
\mathbf{(}N\mathbf{)}$\ and $\mathbf{G}_{(1)}^{\prime }\mathbf{(}N\mathbf{)}$%
\ of Eq.(15) resemble one another except for signs of some elements. As a
result, the second order energies $\mathcal{E}_{(2)}(N)$\ and $\mathcal{E}%
_{(2)}^{\prime }(N)$\ also are uniform and proportional to $N-1$, viz. these
coincide with $\gamma ^{2}(N-1)/2.$ Such a result causes no surprise, as
both linear and cross-conjugated polyenes contain the same numbers of C-C
bonds and thereby of CP(2)s for the same $N$ value.

Let us turn now to elements of the second order matrix $\mathbf{G}_{(2)}$
defined by Eqs.(4) and/or (13). To ensure a non-zero value of the element $%
\mathbf{G}_{(2)il}$\ (and of $\mathbf{G}_{(2)li}$), the underlying BOs $%
\varphi _{(+)i}$ and $\varphi _{(-)l}$\ should belong to second-neighboring
C=C bonds (Sect. 2), i.e. the Ith C=C bond and the Lth one should possess a
common first neighbour coinciding with, say, the Mth C=C bond. The mutual
arrangement of the three involved C=C bonds also playes an important role
here: Given that the whole fragment I-M-L is of a linear constitution (see
e.g. the isomer I of Fig.1), the mediating effect of the BBO $\varphi
_{(+)m} $\ and that of the ABO $\varphi _{(-)m}$\ are added together and,
consequently, a non-zero element $\mathbf{G}_{(2)il}(\mathbf{G}_{(2)li})$\
arises. Meanwhile, the analogous increments cancel out one another for the
cross-conjugated arrangement of C=C bonds I, M and J (e.g. in the isomer IV
of Fig.1), and the relevant element $\mathbf{G}_{(2)il}(\mathbf{G}_{(2)li})$%
\ vanishes. In summary, two non-zero elements ($\mathbf{G}_{(2)il}$ and $%
\mathbf{G}_{(2)li}$) correspond to BOs of the terminal C=C bonds (I and L)
of any linear hexatriene-like fragment I-M-L and thereby to any conjugated
path embracing three C=C bonds and further abbreviated as CP(3) [The
remaining elements $\mathbf{G}_{(2)im}$ and $\mathbf{G}_{(2)ml}$ vanish
because of zero intrabond resonance parameters (see Eq.(7)]. This implies
the total number of non-zero elements of a certain matrix $\mathbf{G}_{(2)}$
to coincide with the two-fold number of CP(3)s in the system under
consideration. Uniform absolute values of the above-specified elements also
easily follow from the definition of Eq.(4). As is seen from Eq.(3), the
matrix $\mathbf{G}_{(2)}$ determines the positive (stabilizing) component ($%
\mathcal{E}_{(4)}^{(+)}$)\ of the fourth order energy that is an additive
function with respect to squares of separate elements $\mathbf{G}_{(2)il}(%
\mathbf{G}_{(2)li})$ in addition. Consequently, the component $\mathcal{E}%
_{(4)}^{(+)}$\ consists of a sum of transferable increments of individual
CP(3)s and thereby it is expected to be proportional to the total number of
these paths.

The above-specified linear and cross-conjugated polyenes may be taken here
again as examples. For the linear isomer, Eqs. (3), (4) and (15) yield the
following common formulae%
\begin{equation}
\mathbf{G}_{(2)}\mathbf{(}N\mathbf{)=}\frac{\gamma ^{2}}{16}\left\vert 
\begin{array}{cccccc}
0 & 0 & 2 & 0 & 0 & .. \\ 
0 & 0 & 0 & 2 & 0 & .. \\ 
-2 & 0 & 0 & 0 & 2 & .. \\ 
0 & -2 & 0 & 0 & 0 & .. \\ 
.. & .. & .. & .. & .. & ..%
\end{array}%
\right\vert ,\quad \mathcal{E}_{(4)}^{(+)}(N)=\frac{8\gamma ^{4}(N-2)}{64},
\end{equation}%
where $\gamma ^{4}/64$ is used here and below as a "subsidiary" unit of the
fourth order energy[15,16]. Elements of the matrix $\mathbf{G}_{(2)}\mathbf{(%
}N\mathbf{)}$\ are chosen to coincide with 2 by choice of the front factor $%
\gamma ^{2}/16$\ instead of $\gamma ^{2}/8$\ [15,16] in order to reflect
participation of mediating orbitals in pairs (e.g. $\varphi _{(+)m}$\ and $%
\varphi _{(-)m}$)\ more conveniently. The expression for $\mathbf{G}_{(2)}%
\mathbf{(}N\mathbf{)}$\ of Eq.(16) illustrates the above-concluded
one-to-one correspondence between non-zero elements of the matrix $\mathbf{G}%
_{(2)}$ and individual CP(3)s. Proportionality between $\mathcal{E}%
_{(4)}^{(+)}(N)$\ and the total number of these paths $(N-2)$ also is seen.
The fact that both $\mathbf{G}_{(2)}\mathbf{(}2\mathbf{)}$ and $\mathcal{E}%
_{(4)}^{(+)}(2)$ vanish for butadiene (N=2) containing no CP(3)s causes no
surprise here. By contrast, the alternating signs of elements when passing
from one line of the matrix $\mathbf{G}_{(1)}^{\prime }(N)$\ of Eq.(15) to
another gives birth to a zero matrix $\mathbf{G}_{(2)}^{\prime }(N)$\ for
dendralenes in accordance with absence of CP(3)s in these hydrocarbons. As a
result, the stabilizing component of the fourth order energy $\mathcal{E}%
_{(4)}^{(+)\prime }(N)$ also vanishes.

Proportionality between the number of non-zero elements of the matrix
$\mathbf{G}_{(2)}$ and that of CP(3)s deserves more illustration.
To this end, let us consider the four isomers of octatetraene I-IV
(Fig.~1). The matrix $\mathbf{G}_{(2)}(I)$\ and the energy increment
$\mathcal{E}_{(4)}^{(+)}(I)$\ of the linear system I result directly from Eq.(16) for
N=4, and $\mathcal{E}_{(4)}^{(+)}(I)$\ equals to $16\gamma ^{4}/64$. The
remaining formulae under our interest are as follows 
\begin{equation*}
\mathbf{G}_{(2)}(II)=\frac{\gamma ^{2}}{16}\left\vert 
\begin{array}{cccc}
0 & 0 & 2 & 2 \\ 
0 & 0 & 0 & 0 \\ 
-2 & 0 & 0 & 0 \\ 
-2 & 0 & 0 & 0%
\end{array}%
\right\vert ,\mathbf{G}_{(2)}(III)=\frac{\gamma ^{2}}{16}\left\vert 
\begin{array}{cccc}
0 & 0 & 0 & 0 \\ 
0 & 0 & 0 & -2 \\ 
0 & 0 & 0 & 0 \\ 
0 & 2 & 0 & 0%
\end{array}%
\right\vert ,\quad \mathbf{G}_{(2)}(IV)=\mathbf{0},
\end{equation*}%
\begin{equation}
\mathcal{E}_{(4)}^{(+)}(II)=\frac{16\gamma ^{4}}{64},\quad \mathcal{E}%
_{(4)}^{(+)}(III)=\frac{8\gamma ^{4}}{64},\quad \mathcal{E}%
_{(4)}^{(+)}(IV)=0.
\end{equation}
Thus, total numbers of non-zero elements of matrices $\mathbf{G}_{(2)}$\
coincide with two-fold numbers of CP(3)s for these systems too, i.e. with 4,
4, 2 and 0 for isomers I-IV, respectively. Moreover, the related energy
components ($\mathcal{E}_{(4)}^{(+)}$) also are proportional to the same
numbers.

Therefore, a simple relation may be concluded between elements of matrices $%
\mathbf{G}_{(1)}$ and $\mathbf{G}_{(2)},$\ on the one hand, and the
conjugated paths CP(2) and CP(3), on the other hand. Moreover, these CPs are
the only conjugated fragments participating in the formation of elements
concerned. Additivity of the consequent energetic increments $\mathcal{E}%
_{(2)}$\ and $\mathcal{E}_{(4)}^{(+)}$ with respect to transferable
contributions of CP(2)s and CP(3)s, respectively, also is among conclusions
here.

Such a simple state of things, however, is no longer preserved when passing
to terms of higher orders. To demonstrate this, we are about to consider
elements of the matrix $\mathbf{G}_{(3)}^{o}$ separately.

\subsection{Analysis of elements of the third order matrix $G_{(3)}^{o}$}

Let us start with the above-discussed linear polyene containing $N-3$
CP(4)s, where N$\geq 3.$ The respective common third order matrix $\mathbf{G}%
_{(3)}^{o}(N)$\ and the related sixth order energy increment $\mathcal{E}%
_{(6)1}^{(+)}(N)$\ follow from Eqs.(8), (12) and (16), viz.%
\begin{equation}
\mathbf{G}_{(3)}^{o}(N)\mathbf{=-}\frac{\gamma ^{3}}{32}\left\vert 
\begin{array}{ccccccc}
0 & 1 & 0 & 2 & 0 & 0 & ... \\ 
-1 & 0 & 2 & 0 & 2 & 0 & ... \\ 
0 & -2 & 0 & 2 & 0 & 2 & ... \\ 
-2 & 0 & -2 & 0 & 2 & 0 & ... \\ 
0 & -2 & 0 & -2 & 0 & 2 & ... \\ 
.. & .. & .. & .. & .. & .. & ..%
\end{array}%
\right\vert ,\quad \mathcal{E}_{(6)1}^{(+)}(N)=\frac{4\gamma ^{6}[4(N-3)+1]}{%
256}.
\end{equation}%
where $-$\ $\gamma ^{3}/32$ serves here and below as the standard factor for
matrices $\mathbf{G}_{(3)}^{o}$. Accordingly, $\gamma ^{6}/256$\ will be
used as the "subsidiary" sixth order energy unit. The first representatives
of the series of matrices $\mathbf{G}_{(3)}^{o}(N)$ and of energy increments 
$\mathcal{E}_{(6)1}^{(+)}(N)$\ corresponding to N=3 and 4 also deserve
exhibiting, viz.%
\begin{eqnarray}
\mathbf{G}_{(3)}^{o}(3) &=&-\frac{\gamma ^{3}}{32}\left\vert 
\begin{array}{ccc}
0 & 1 & 0 \\ 
-1 & 0 & 1 \\ 
0 & -1 & 0%
\end{array}%
\right\vert ,\quad \mathbf{G}_{(3)}^{o}(4)\equiv \mathbf{G}_{(3)}^{o}(I)=-%
\frac{\gamma ^{3}}{32}\left\vert 
\begin{array}{cccc}
0 & 1 & 0 & 2 \\ 
-1 & 0 & 2 & 0 \\ 
0 & -2 & 0 & 1 \\ 
-2 & 0 & -1 & 0%
\end{array}%
\right\vert ,  \notag \\
\mathcal{E}_{(6)1}^{(+)}(3) &=&\frac{4\gamma ^{6}}{256},\quad \mathcal{E}%
_{(6)1}^{(+)}(4)\equiv \mathcal{E}_{(6)1}^{(+)}(I)=\frac{20\gamma ^{6}}{256}.
\end{eqnarray}
These particular cases evidently represent linear isomers of hexatriene and
of octatetraene I, respectively [Note that the energy increment
$\mathcal{E}_{(6)1}^{(+)}(3)$ follows directly from Eq.(18) after substituting N=3, but
it is not the case for $\mathbf{G}_{(3)}^{o}(3).$ A more detailed
discussion of this point may be found in Ref.[15]]. As is seen from Eq.(18),
the energy increment $\mathcal{E}_{(6)1}^{(+)}(N)$ contains a dependence
upon the number of CP(4)s of the given chain ($N-3$) in accordance with the
expectation. The total number of non-zero elements of the matrix
$\mathbf{G}_{(3)}^{o}(N),$\ however, exceeds the two-fold number of CP(4)s
considerably. Moreover, significant elements correspond not only to BOs of
third-neighbouring C=C bonds (as it may be expected on the basis of the
above experience), but also to orbitals of their first-neighbouring pairs.
For example, the matrix $\mathbf{G}_{(3)}^{o}(4)[\mathbf{G}_{(3)}^{o}(I)]$
contains eight non-zero elements referring to BOs of C=C bonds under numbers
(1,2), (1,4), (2,3) and (3,4) (Fig.1), and the elements concerned take
non-uniform values in addition, e. g. 2 and 1 for pairs of BOs $(+)1,(-)4$
and $(+)1,(-)2$, respectively. It is evident that all the above-enumerated
elements contribute to the stabilizing increment $\mathcal{E}_{(6)1}^{(+)}(N)$.
Finally, neither the matrix $\mathbf{G}_{(3)}^{o}(3)$
itself nor the relevant energy correction $\mathcal{E}_{(6)1}^{(+)}(3)$
vanish for the three-membered system of hexatriene ($N=3$) containing no
CP(4)s [in contrast to the zero matrix $\mathbf{G}_{(2)}(2)$ and the
vanishing correction $\mathcal{E}_{(4)}^{(+)}(2)$ of butadiene ($N=2$)
discussed in the previous subsection].

To clarify the origin of these distinctive results, let us write down
explicit expressions for elements of the matrix $\mathbf{G}_{(3)}^{o}(4)[%
\mathbf{G}_{(3)}^{o}(I)]$. Using the first relation of Eq.(12), we obtain%
\begin{eqnarray}
\mathbf{G}_{(3)14}^{o}(I) &=&-\frac{1}{2}(\mathbf{S}_{12}\mathbf{G}_{(2)24}-%
\mathbf{G}_{(2)13}\mathbf{Q}_{34}),  \notag \\
\mathbf{G}_{(3)23}^{o}(I) &=&-\frac{1}{2}(\mathbf{S}_{21}\mathbf{G}_{(2)13}-%
\mathbf{G}_{(2)24}\mathbf{Q}_{43}),  \notag \\
\mathbf{G}_{(3)12}^{o}(I) &=&\frac{1}{2}\mathbf{G}_{(2)13}\mathbf{Q}%
_{32},\quad \mathbf{G}_{(3)34}^{o}(I)=-\frac{1}{2}\mathbf{S}_{32}\mathbf{G}%
_{(2)24}.
\end{eqnarray}%
As is seen from the first formula of Eq.(20), a simple linear pathway from
the BBO of the bond C$_{1}$=C$_{5}$ ($\varphi _{(+)1}$) to the ABO of the C$%
_{4}$=C$_{8}$ ($\varphi _{(-)4}$) underlies the element $\mathbf{G}%
_{(3)14}^{o}(I).$\ Moreover, mediating effects of intervening orbitals ($%
\varphi _{(+)2}$ and $\varphi _{(-)3}$) may be easily shown to be added
together here. As a result, the absolute value of the element concerned
coincides with 2. Thus, a simple relation may be concluded immediately
between the element $\mathbf{G}_{(3)14}^{o}(I)$\ and the only CP(4) of the
system I. A similar addition of contributing components takes place in the
expression for the element $\mathbf{G}_{(3)23}^{o}(I)$\ too and the
resulting value coincides with that of $\mathbf{G}_{(3)14}^{o}(I)$. The
underlying pathways, however, differ from linear ones in the latter case.
Indeed, pathways over BOs containing self-returning segments correspond to
both components of the expression for $\mathbf{G}_{(3)23}^{o}(I)$, wherein
orbitals of terminal bonds ($\varphi _{(+)1}$ and $\varphi _{(-)4}$)
participate as mediators. Meanwhile, orbitals of the fourth (C$_{4}$=C$_{8}$%
) and first (C$_{1}$=C$_{5}$) bonds play no role in the formation of the
first and second component of the element $\mathbf{G}_{(3)23}^{o}(I),$\
respectively. It is evident that conjugated paths CP(3) embracing triplets
of C=C bonds under numbers 1,2,3 and 2,3,4 may be correspondingly ascribed
to the above-specified components. The same refers also to elements $\mathbf{%
G}_{(3)12}^{o}(I)$\ and $\mathbf{G}_{(3)34}^{o}(I).$

We may expect, therefore, that matrices $\mathbf{G}_{(3)}^{o}$\ generally
contain information not only about CP(4)s of the given polyene, but also
about shorter conjugated paths. It is also likely that the presence of a
standard CP(4) is not among necessary conditions for a non-zero matrix $%
\mathbf{G}_{(3)}^{o}$\ to represent a certain pi-electron system. To support
these anticipations, let us consider some polyenes of more involved
constitutions.

Let us start with the isomer of octatetraene III (Fig.~1) containing both
conjugated and cross-conjugated fragments. As opposed to its linear
counterpart I, the new isomer III contains no CP(4). Nevertheless, it is
characterized by a non-zero matrix $\mathbf{G}_{(3)}^{o}(III)$ and a
significant energy increment $\mathcal{E}_{(6)1}^{(+)}(III)$, viz.
\begin{equation}
\mathbf{G}_{(3)}^{o}(III)=-\frac{\gamma ^{3}}{32}\left\vert 
\begin{array}{cccc}
0 & 0 & 0 & -1 \\ 
0 & 0 & -1 & 0 \\ 
0 & 1 & 0 & -1 \\ 
1 & 0 & 1 & 0%
\end{array}%
\right\vert ,\quad \mathcal{E}_{(6)1}^{(+)}(III)=\frac{6\gamma ^{6}}{256}.
\end{equation}%
Moreover, elements $\mathbf{G}_{(3)14}^{o}$\ and $\mathbf{G}_{(3)41}^{o}$\
take non-zero values in the matrix $\mathbf{G}_{(3)}^{o}(III)$\ in spite of
the fact that the terminal C=C bonds (C$_{1}$=C$_{5}$ and C$_{4}$=C$_{8}$)
are not joined by a conjugated path. This result causes little surprise if
we recall the definition of the matrix $\mathbf{G}_{(3)}^{o}$\ in terms of $%
\mathbf{G}_{(2)}$\ shown in Eq.(12). Indeed, this definition indicates
non-zero values of elements $\mathbf{S}_{12}$ and $\mathbf{G}_{(2)24}$\ to
be sufficient to ensure a significant element $\mathbf{G}_{(3)14}^{o}$\ and
this condition is met by the terminal orbitals $\varphi _{(+)1}$ and $%
\varphi _{(-)4}$\ of our system III. Again, some similarity is beyond any
doubt between constitutions of matrices $\mathbf{G}_{(3)}^{o}(I)$ of Eq.(19)
and $\mathbf{G}_{(3)}^{o}(III)$ of Eq.(21). These two points allow the
isomer III to be considered as a partially conjugated system. In this
connection, a new concept of the semi-conjugated path may be introduced that
contributes to the sixth order stabilization of the system along with the
usual CP(4)s. In the present case, we will have to do with a semi-conjugated
path embracing four C=C bonds and abbreviated below by SCP(4). The lower
stabilizing effect of this new path as compared to the standard increment of
the only CP(4) of the linear isomer I (see Eq.(19)) causes no surprise here.

Another important distinction between the third order matrices $\mathbf{G}%
_{(3)}^{o}$ along with the related energy increments $\mathcal{E}%
_{(6)1}^{(+)}$ and their counterparts of lower orders (Subsect. 3.1)
consists in much more involved dependences of the third (sixth) order
characteristics upon the numbers of the standard conjugated paths when
passing from linear to branched polyenes. This distinction may be traced
back to the fact that the side subchains of branched systems offer new
self-returning segments for pathways underlying separate elements $\mathbf{G}%
_{(3)il}^{o}(\mathbf{G}_{(3)li}^{o})$ and thereby the absolute values of
these elements become excessively increased. For example, the branched
octatetraene II is characterized by the following matrix $\mathbf{G}%
_{(3)}^{o}(II)$\ and the energy increment $\mathcal{E}_{(6)1}^{(+)}(II):$ 
\begin{equation}
\mathbf{G}_{(3)}^{o}(II)=-\frac{\gamma ^{3}}{32}\left\vert 
\begin{array}{cccc}
0 & 2 & 0 & 0 \\ 
-2 & 0 & 1 & 1 \\ 
0 & -1 & 0 & 0 \\ 
0 & -1 & 0 & 0%
\end{array}%
\right\vert ,\quad \mathcal{E}_{(6)1}^{(+)}(II)=\frac{12\gamma ^{6}}{256}.
\end{equation}%
It is seen that absolute values of elements $\mathbf{G}_{(3)12}^{o}(II)$\
and $\mathbf{G}_{(3)21}^{o}(II)$\ coincide with $\mathbf{G}_{(3)14}^{o}(I)$\
and $\mathbf{G}_{(3)23}^{o}(I)$\ of Eq.(19) in spite of absence of CP(4)s in
the branched system II. These increased elements are unambiguosly related to
emergence of two self-returning segments in the pathways over BOs underlying
elements $\mathbf{G}_{(3)12}^{o}(II)$\ and $\mathbf{G}_{(3)21}^{o}(II),$\
namely of segments embracing the bonds C$_{4}$=C$_{8}$ and C$_{3}$=C$_{7}$.
Nevertheless, the total value of the stabilizing increment $\mathcal{E}%
_{(6)1}^{(+)}(II)$\ is almost two times smaller for the branched isomer II
as compared to the relevant value ($20\gamma ^{6}/256$) for its linear
counterpart I.

\begin{figure}
\includegraphics[width=0.9\textwidth]{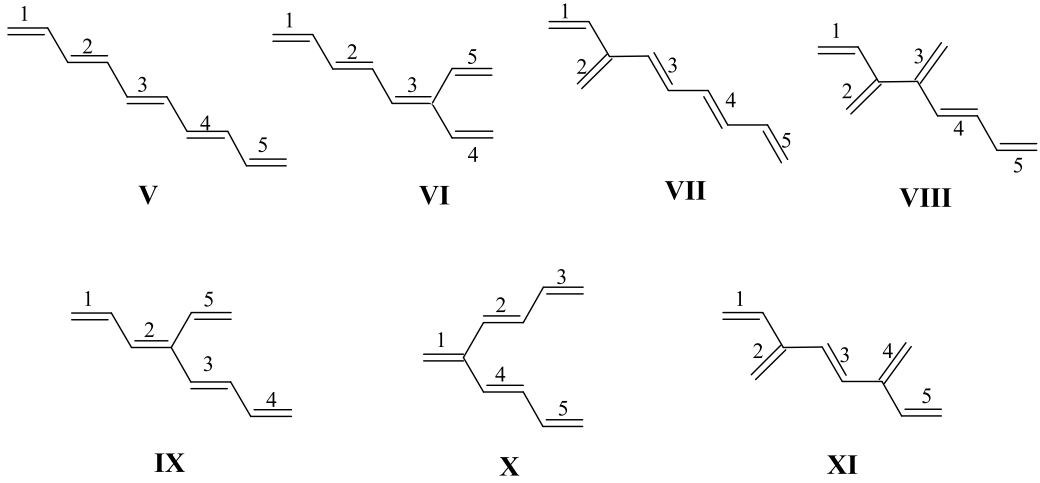}
\caption{Isomers of decapentaene (V-XI) containing five C=C bonds (N=5).
Numberings of these bonds also are shown}
\end{figure}

The decisive role of the side subchains in the formation of matrices
$\mathbf{G}_{(3)}^{o}$\ (and thereby of energy components
$\mathcal{E}_{(6)1}^{(+)}$) may be further illustrated by comparing these
characteristics for isomers of decapentaene V, VI and IX (Fig. 2), where
N=5. The total numbers of CP(4)s correspondingly equal to 2, 2 and 1 in
these systems. Again, the isomers concerned may be regarded as consisting of
the principal linear chain and of a side subchain, embracing the C=C bonds
under numbers 1-4 and 5, respectively. The side subchain (C$_{5}$=C$_{10}$)
then takes distinct positions with respect to the principal chain in the
systems under comparison and thereby it offers different sets of
self-returning segments for pathways over BOs underlying particular elements
of matrices $\mathbf{G}_{(3)}^{o}(V),$ $\mathbf{G}_{(3)}^{o}(VI)$ and
$\mathbf{G}_{(3)}^{o}(IX)$. As a result, distinctions may be anticipated both
in the constitutions of the above-enumerated matrices and in values of the
consequent energy components.

To demonstrate this, let us start with matrices $\mathbf{G}_{(3)}^{o}$. The
matrix $\mathbf{G}_{(3)}^{o}(V)[\mathbf{G}_{(3)}^{o}(5)]$ easily results
from the general expression of Eq.(18) and contains eight elements equal to
either 2 or -2 and four elements coinciding with either 1 or -1. Meanwhile,
the remaining matrices and the consequent energy increments are as follows
\begin{align}
\mathbf{G}_{(3)}^{o}(VI) =&-\frac{\gamma ^{3}}{32}\left\vert 
\begin{array}{ccccc}
0 & 1 & 0 & 2 & 2 \\ 
-1 & 0 & 3 & 0 & 0 \\ 
0 & -3 & 0 & 1 & 1 \\ 
-2 & 0 & -1 & 0 & 0 \\ 
-2 & 0 & -1 & 0 & 0
\end{array}
\right\vert ,\mathbf{G}_{(3)}^{o}(IX)=-\frac{\gamma ^{3}}{32}\left\vert 
\begin{array}{ccccc}
0 & 2 & 0 & 2 & 0 \\ 
-2 & 0 & 2 & 0 & 1 \\ 
0 & -2 & 0 & 1 & 0 \\ 
-2 & 0 & -1 & 0 & -1 \\ 
0 & -1 & 0 & 1 & 0
\end{array}
\right\vert ,  \notag \\
\mathcal{E}_{(6)1}^{(+)}(V) =&\frac{36\gamma ^{6}}{256},\quad
\mathcal{E}_{(6)1}^{(+)}(VI)=\frac{40\gamma ^{6}}{256},\quad
\mathcal{E}_{(6)1}^{(+)}(IX)=\frac{30\gamma ^{6}}{256}.
\end{align}
It is seen that the matrix $\mathbf{G}_{(3)}^{o}(VI)$\ contains elements of
higher absolute values as compared to $\mathbf{G}_{(3)}^{o}(IX)$ and $%
\mathcal{E}_{(6)1}^{(+)}(VI)$ accordingly exceeds $\mathcal{E}%
_{(6)1}^{(+)}(IX)$. This result is in line with distinct numbers of CP(4)s
in the isomers VI and IX (2 and 1) and thereby causes little surprise. It
deserves adding, however, that the energy component $\mathcal{E}%
_{(6)1}^{(+)}(IX)$ does not coincide with $\mathcal{E}_{(6)1}^{(+)}(I)$\ of
Eq.(19) in spite of the same number of CP(4)s in the systems I and IX (equal
to 1). Comparison of matrices $\mathbf{G}_{(3)}^{o}(V)$\ and $\mathbf{G}%
_{(3)}^{o}(VI)$ yields even more unexpected conclusions. Indeed, the matrix $%
\mathbf{G}_{(3)}^{o}(VI)$\ differs from $\mathbf{G}_{(3)}^{o}(V)$\
significantly in spite of the same number of CP(4)s present in both systems.
Moreover, the sum of squares of elements of the former matrix exceeds that
of the latter and, consequently, the branched isomer VI proves to be
described by a higher energy increment $\mathcal{E}_{(6)1}^{(+)}$\ as
compared to its linear counterpart (V). This result implies a dependence of
the sixth order energy upon the mutual arrangement of the two CP(4)s [Note
that an increased overall stability of the branched system VI vs. the linear
one (V) is not among the implications as discussed in Sect. 4].

In summary, a relation is beyond any doubt between the third order matrix
$\mathbf{G}_{(3)}^{o}$ representing a certain polyene and the number of
CP(4)s present there as it was the case with matrices $\mathbf{G}_{(1)}$
and $\mathbf{G}_{(2)}$ determined by numbers of CP(2)s and CP(3)s,
respectively (Subsect. 3.1). In contrast to the latter cases, however, the
total number of CP(4)s is not the only factor determining the given matrix
$\mathbf{G}_{(3)}^{o}$\ and thereby the consequent energy increment
$\mathcal{E}_{(6)1}^{(+)}$: Other details of constitution of the given system also
play their role here, e.g. presence of semi-conjugated paths (SCP(4)s) and a
particular mutual arrangement of several CP(4)s (if any).

Let us turn now to elements of matrix products of Eqs.(3) and (9)--(11).

\subsection{The role of composite conjugated paths in the formation of
elements of matrix products }

Let us start with the product $\mathbf{G}_{(1)}\mathbf{G}_{(1)}^{+}$\
determining the fourth order destabilizing energy component $\mathcal{E}%
_{(4)}^{(-)}$\ of Eq.(3). It is evident that an element $(\mathbf{G}_{(1)}%
\mathbf{G}_{(1)}^{+})_{ij}$ does not vanish if in the given system there is
an ABO $\varphi _{(-)l}$\ such that $\mathbf{G}_{(1)il}\neq 0$ and $\mathbf{G%
}_{(1)lj}^{+}\equiv \mathbf{G}_{(1)jl}\neq 0.$\ Because of the equality $%
\mathbf{G}_{(1)ii}=0$\ for any $i$ (see Eq.(7)), ABOs $\varphi _{(-)i}$ and $%
\varphi _{(-)j}$\ are not able to play this role. Thus, the ABO $\varphi
_{(-)l}$\ necessarily belongs to a third (say Lth) C=C bond, where L$\neq $%
I\ and L$\neq $J. In other words, two simple mutually connected pathways are
required here, namely a pathway from $\varphi _{(+)i}$ \ to $\varphi _{(-)l}$%
\ and that from $\varphi _{(-)l}$ \ and $\varphi _{(+)j}.$ Since a CP(2)
corresponds to any element $\mathbf{G}_{(1)il}$ (Subsect. 3.1), the above
condition resolves itself into a requirement of two simple CP(2)s embracing
a common (Lth) C=C bond. Given that this is the case, we will say that in
the given system there is a composite conjugated path over three C=C bonds
abbreviated below by CCP(3). Besides, a CCP(3) is automatically ensured
under presence of a standard CP(3), but not vice versa (see Sect. 4).
Further, the matrix product $\mathbf{G}_{(1)}\mathbf{G}_{(1)}^{+}$\ is
characterized by non-zero diagonal elements $(\mathbf{G}_{(1)}\mathbf{G}%
_{(1)}^{+})_{ii}$\ that are interpretable as indirect self-interactions of
respective BBOs $\varphi _{(+)i}$ \ via ABOs of the neighbouring C=C bonds
(Sect. 2). In this connection, let us also define self-returning composite
conjugated paths coinciding with squares of CP(2)s. As with the latter,
these new paths also embrace pairs of neighbouring C=C bonds. Thus, let us
use the abbreviation SRCCP(2). On the whole, the matrix $\mathbf{G}_{(1)}%
\mathbf{G}_{(1)}^{+}$\ of a certain polyene may be then expected to contain
information about both CCP(3)s and SRCCP(2)s.

For illustration, let us consider matrices $\mathbf{G}_{(1)}\mathbf{G}_{(1)}^{+}$
representing the linear octatetraene (I), as well as its
branched and cross-conjugated isomers II and IV (Fig. 1). The first two
matrices under comparison differ one from another significantly, especially
in respect of diagonal elements, viz.
\begin{equation}
\mathbf{G}_{(1)}\mathbf{G}_{(1)}^{+}(I)=\frac{\gamma ^{2}}{16}\left\vert 
\begin{array}{cccc}
1 & 0 & -1 & 0 \\ 
0 & 2 & 0 & -1 \\ 
-1 & 0 & 2 & 0 \\ 
0 & -1 & 0 & 1
\end{array}
\right\vert ,\mathbf{G}_{(1)}\mathbf{G}_{(1)}^{+}(II)=\frac{\gamma ^{2}}{16}
\left\vert 
\begin{array}{cccc}
1 & 0 & -1 & -1 \\ 
0 & 3 & 0 & 0 \\ 
-1 & 0 & 1 & 1 \\ 
-1 & 0 & 1 & 1
\end{array}
\right\vert .
\end{equation}
The reason for this distinction consists in the increased number of first
neighbours of the second (C$_{2}$=C$_{6}$) bond in the branched isomer II
and thereby in the larger indirect self-interaction of the relevant BBO $%
\varphi _{(+)2}$. In terms of conjugated paths we have to do here with an
increased number of SRCCP(2)s referring to the 2nd C=C bond. If we recall
here that isomers I and II both contain two CP(3)s, the above result becomes
even more important in distinguishing between their stabilities (Sect. 4).
So far as the matrix $\mathbf{G}_{(1)}\mathbf{G}_{(1)}^{+}(IV)$ is
concerned, it resembles $\mathbf{G}_{(1)}\mathbf{G}_{(1)}^{+}(I)$ in
respect of absolute values of all matrix elements [negative off-diagonal
elements of $\mathbf{G}_{(1)}\mathbf{G}_{(1)}^{+}(I)$ become replaced by
positive ones when passing to $\mathbf{G}_{(1)}\mathbf{G}_{(1)}^{+}(IV)$].
This implies the linear isomer I and the cross-conjugated one IV to contain
the same sets of both linear and self-returning composite conjugated paths
(i.e. of CCP(3)s and SRCCP(2)s) in spite of different numbers of the
standard CP(3)s (2 and 0).

Let us turn now to third order matrix products and start with $\mathbf{G}%
_{(1)}\mathbf{G}_{(1)}^{+}\mathbf{G}_{(1)}\mathbf{.}$ The skew-symmetric
(skew-Hermitian)\ nature of this matrix (Sect. 2) implies zero values for
any diagonal element $(\mathbf{G}_{(1)}\mathbf{G}_{(1)}^{+}\mathbf{G}%
_{(1)})_{ii}.$\ Nevertheless, an analogy still exists between products $%
\mathbf{G}_{(1)}\mathbf{G}_{(1)}^{+}\mathbf{G}_{(1)}$\ and $\mathbf{G}_{(1)}%
\mathbf{G}_{(1)}^{+}$. In particular, the matrix $\mathbf{G}_{(1)}\mathbf{G}%
_{(1)}^{+}\mathbf{G}_{(1)}$\ may be similarly shown to represent products of
three connected CP(2)s, the total number of the embraced C=C bonds generally
coinciding with four. Accordingly, we may define composite conjugated paths
over four C=C bonds (CCP(4)s). It should be mentioned, however, that the
overall situation becomes somewhat more involved when passing from $\mathbf{G%
}_{(1)}\mathbf{G}_{(1)}^{+}$\ to $\mathbf{G}_{(1)}\mathbf{G}_{(1)}^{+}%
\mathbf{G}_{(1)},$\ as it was the case when comparing matrices $\mathbf{G}%
_{(2)}$ and $\mathbf{G}_{(3)}^{o}$\ (Sect. 3). In particular, pathways over
BOs underlying elements $(\mathbf{G}_{(1)}\mathbf{G}_{(1)}^{+}\mathbf{G}%
_{(1)})_{il}$\ may possess self-returning segments (as it was the case with $%
\mathbf{G}_{(3)il}^{o}$)\textbf{.}\ Consequently, the resulting CCPs
actually embrace three or even two C=C bonds. It is also evident that zero
diagonal elements of the product $\mathbf{G}_{(1)}\mathbf{G}_{(1)}^{+}%
\mathbf{G}_{(1)}$\ reflect impossibility of completely self-returning paths
in this case. Finally, matrix products $\mathbf{G}_{(1)}\mathbf{G}_{(2)}^{+}$%
\ and $\mathbf{G}_{(2)}\mathbf{G}_{(1)}^{+}$\ remain to be discussed. In
contrast to previous cases, these products are neither symmetric (Hermitian)
nor skew-symmetric (skew-Hermitian) matrices. It is evident that the
underlying CCPs consist of products of a CP(2) and a CP(3), and of a CP(3)
and a CP(2), respectively, and also generally embrace four C=C bonds.
Self-returning segments are possible here too.

It is seen, therefore, that elements of matrix products are determined by
conjugated paths of a non-standard (viz. composite) nature. Moreover, most
of these new paths give birth to destabilizing energy components (e.g. $%
\mathcal{E}_{(4)}^{(-)}$\ and $\mathcal{E}_{(6)}^{(-)}$).

Before finishing this Section, the following remark deserves to be made:
Energy increments originating from matrix products $\mathbf{G}_{(1)}\mathbf{G%
}_{(1)}^{+},$ $\mathbf{G}_{(1)}\mathbf{G}_{(2)}^{+},$ etc. and these related
to simple matrices $\mathbf{G}_{(1)},\mathbf{G}_{(2)}$ and $\mathbf{G}%
_{(3)}^{o}$\ hardly are independent. Quite the reverse, a certain
interdependence may be foreseen between some of these increments, e.g.
between $\mathcal{E}_{(6)1}^{(+)}$\ and $\mathcal{E}_{(6)}^{(-)}.$\ The main
reason for such an anticipation consists in the presence of the same matrix $%
\mathbf{G}_{(2)}$\ in the definitions of underlying matrices $\mathbf{G}%
_{(3)}^{o}$ and $\mathbf{G}_{(1)}\mathbf{G}_{(2)}^{+}$\ (see Eq.(12)).
[Besides, both matrices $\mathbf{G}_{(3)}^{o}$ and $\mathbf{G}_{(1)}\mathbf{G%
}_{(2)}^{+}$ vanish, if $\mathbf{G}_{(2)}$\ coincides with a zero matrix].
Thus, an increased stabilization of a certain system due to a larger
increment $\mathcal{E}_{(6)1}^{(+)}$\ may be expected to be accompanied by a
growing destabilization ($\mathcal{E}_{(6)}^{(-)}$) and vice versa. Given
that relative stabilities of different isomers are under interest (Sect. 4),
the outcome of the comparison then depends on the overall balance between
energy increments of opposite signs.

\vspace{4mm}

\section{Discussions of relative stabilities of specific isomers}

Let us start with comparison of linear polyenes to their cross-conjugated
isomers (dendralenes) in respect of overall relative stabilities of their
pi-electron systems. Zero order energies of these isomers are uniform for
the same number of C=C bonds (N), viz. $\mathcal{E}_{(0)}^{{}}(N)=$ $%
\mathcal{E}_{(0)}^{\prime }(N)=2N$. The relevant second order increments
also coincide one with another (see the implications of Eq.(15)). Finally,
the stabilizing components of the fourth order energies are discussed in the
Subsect. 3.1 [$\mathcal{E}_{(4)}^{(+)}(N)$ is shown in Eq.(16) and takes a
significant value because of N-2 three-membered conjugated paths (CP(3)s)
present in the linear system, whereas $\mathcal{E}_{(4)}^{(+)\prime }(N)$\
vanishes due to absence of these paths in dendralenes]. Thus, let us turn
immediately to the remaining (destabilizing) components of the fourth order
energies.

Employment of the expression for the matrix $\mathbf{G}_{(1)}(N)$\ of
Eq.(15) to construct the product $(\mathbf{G}_{(1)}\mathbf{G}_{(1)}^{+})(N)$%
\ shows this important representation of linear polyenes to contain elements
1,2,2...2,1 in its principal diagonale along with $-$1 in the
second-neighbouring off-diagonal positions [The matrix $(\mathbf{G}_{(1)}%
\mathbf{G}_{(1)}^{+})(I)$\ of Eq.(24)\ serves as an example for N=4].
Moreover, the analogous matrix product $(\mathbf{G}_{(1)}\mathbf{G}%
_{(1)}^{+})^{\prime }(N)$\ representing the dendralene series and
originating from $\mathbf{G}_{(1)}^{\prime }\mathbf{(}N\mathbf{)}$\ of
Eq.(15) also closely resembles the above-discussed one, except for opposite
(i.e. positive) signs of all off-diagonal elements. Accordingly,\
destabilizing components of fourth order energies are uniform for both
systems under comparison. The latter result may be entirely traced back to
an evident fact that products of CP(2)s and thereby numbers of composite
conjugated paths CCP(3)s coincide one with another in both polyenes. The
relevant total fourth order energies are then as follows 
\begin{equation}
\mathcal{E}_{(4)}(N)=\frac{2\gamma ^{4}(N-3)}{64},\;\mathcal{E}%
_{(4)}^{\prime }(N)=\mathcal{E}_{(4)}^{(-)\prime }(N)=\mathcal{E}%
_{(4)}^{(-)}(N)=-\frac{\gamma ^{4}[6(N-2)+2]}{64},
\end{equation}%
where $\mathcal{E}_{(4)}^{(+)}(N)$\ is taken from Eq.(16). It is seen that
the correction $\mathcal{E}_{(4)}(N)$ is a positive quantity for N$\geqslant 
$4 owing to predominance of its stabilizing component $\mathcal{E}%
_{(4)}^{(+)}(N)$ over the destabilizing one $\mathcal{E}_{(4)}^{(-)}(N)$.
This implies the sufficiently long linear polyenes to be additionally
stabilized vs. the sum $\mathcal{E}_{(0)}(N)+\mathcal{E}_{(2)}(N)$\ due to
the fourth order energy. By contrast, the analogous correction of
dendralenes ($\mathcal{E}_{(4)}^{\prime }(N)$) consists of the destabilizing
component only and, consequently, it is a negative quantity [Particular
cases of Eq.(25) referring to N=2 and N=3 also are of interest. In the case
of butadiene (N=2), both relations yield the same correction $\mathcal{E}%
_{(4)}(2)$ equal to $-2\gamma ^{4}/64$\ in accordance with the expectation,
which is a negative quantity in addition due to absence of CP(3)s. This
correction consists of the destabilizing component only that originates from
SRCCP(2)s. For the linear hexatriene (N=3), the two components ($\mathcal{E}%
_{(4)}^{(+)}(3)$ and $\mathcal{E}_{(4)}^{(-)}(3)$) cancel out one another
and the total fourth order energy takes a zero value. The branched isomer of
hexatriene, in turn, is characterized by a negative fourth order energy].

Thus, a higher relative stability of linear polyenes vs. dendralenes
unambiguosly follows from our results and this conclusion coincides with
those of other approaches [32-35]. As with the standard model of conjugated
paths [5], the above analysis also indicates the presence of CP(3)s to be
the origin of the increased stability of linear isomers. In contrast to the
standard model, however, an additional destabilizing factor is now revealed
to manifest itself in both systems under comparison that is interpretable as
a contribution of composite conjugated paths. Only because of the
above-established coincidence of absolute values of the underlying energy
increments for linear and cross-conjugated systems, the destabilizing factor
becomes irrelevant when comparing their relative stabilities.

For other series of polyenes (e.g. the branched ones), a general analysis
like that carried out above hardly is feasible. Thus, we will confine
ourselves to comparisons of relative stabilities of individual
representatives of different series.

Let us start with the four isomers of octatetraene I-IV (Fig.1) containing
the same number of C-C bonds and thereby of CP(2)s (equal to 3).
Accordingly, the relevant second order energies also are uniform. Again, the
total numbers of CP(3)s correspondingly coincide with 2, 2, 1 and 0 for
systems I-IV. Since the linear isomer (I) and its cross-conjugated analogue
(IV) are particular cases of the above-considered polyenes, Eq.(25) yields $%
\mathcal{E}_{(4)}(I)$\ and $\mathcal{E}_{(4)}(IV)$\ equal to $2\gamma ^{4}/64
$ and $-14\gamma ^{4}/64,$ respectively.

Let us now dwell on the branched isomer II. The stabilizing component of the
fourth order energy $\mathcal{E}_{(4)}^{(+)}(II)$ is shown in Eq.(17) and
coincides with $\mathcal{E}_{(4)}^{(+)}(I)$\ following from Eq.(16) for N=4
owing to similar non-zero elements of matrices $\mathbf{G}_{(2)}(II)$\ and $%
\mathbf{G}_{(2)}(I)$ (Subsect. 3.1). This result is in line with the same
number of CP(3)s in polyenes I and II. Meanwhile, the matrix $\mathbf{G}%
_{(1)}\mathbf{G}_{(1)}^{+}(II)$\ differs from both $\mathbf{G}_{(1)}\mathbf{G%
}_{(1)}^{+}(I)$ and $\mathbf{G}_{(1)}\mathbf{G}_{(1)}^{+}(IV)$\
significantly (see Eq.(24) \ and the discussion nearby). It is evident that
the sum of squares of elements of the matrix $\mathbf{G}_{(1)}\mathbf{G}%
_{(1)}^{+}(II)$\ exceeds the relevant value for isomers I and IV. As a
result, the destabilizing component $\mathcal{E}_{(4)}^{(-)}(II)$\ is of an
increased absolute value vs. $\mathcal{E}_{(4)}^{(-)}(I)$\ and $\mathcal{E}%
_{(4)}^{(-)}(IV)$.\ The overall result referring to the isomer II is then as
follows%
\begin{equation}
\mathcal{E}_{(4)}^{(+)}(II)=\frac{16\gamma ^{4}}{64},\quad \mathcal{E}%
_{(4)}^{(-)}(II)=-\frac{18\gamma ^{4}}{64},\quad \mathcal{E}_{(4)}(II)=-%
\frac{2\gamma ^{4}}{64}.
\end{equation}%
Hence, the total fourth order energies $\mathcal{E}_{(4)}(I)$\ and $\mathcal{%
E}_{(4)}(II)$\ differ one from another for isomers I and II in spite of the
same numbers of CP(3)s. Moreover, the linear isomer is predicted to be more
stable as compared to the branched one (II). Although this result is in line
with predictions of the standard CP model (as well as with the relevant
general graph-theoretical results [34,35]), the above analysis indicates
another underlying reason. Indeed, the branched isomer II is now concluded
to be less stable owing to a greater destabilizing effect of self-returning
composite conjugated paths (SRCCP(2)s) defined in Sect. 3. Meanwhile, the
greater stability of the isomer I (vs. the branched analogue II) is traced
back to the presence of a CP(4) in the linear chain when the usual model of
conjugated paths is applied.

Finally, the semi-conjugated isomer (III) remains to be discussed. The
relevant matrix $\mathbf{G}_{(2)}(III)$\ is shown in Eq.(17) and contains
two non-zero elements in accordance with a single CP(3) present in the given
system. Meanwhile, the matrix $\mathbf{G}_{(1)}\mathbf{G}_{(1)}^{+}(III)$\
closely resembles $\mathbf{G}_{(1)}\mathbf{G}_{(1)}^{+}(I)$\ of Eq.(24) in
respect of absolute values of non-zero elements. We then obtain 
\begin{equation}
\mathcal{E}_{(4)}^{(+)}(III)=\frac{8\gamma ^{4}}{64},\quad \mathcal{E}%
_{(4)}^{(-)}(III)=-\frac{14\gamma ^{4}}{64},\quad \mathcal{E}_{(4)}(III)=-%
\frac{6\gamma ^{4}}{64}.
\end{equation}
Thus, the above results indicate the following order of relative stabilities
of isomers: $I>II>III>IV$. Completely
similar conclusions follow also for analogous isomers of decapentaene V, VI,
VII and VIII shown in Fig. 2. The relevant fourth order energies
correspondingly coincide with $4\gamma ^{4}/64,$\ $0,$ $-4\gamma ^{4}/64$
and $-12\gamma ^{4}/64.$Besides, the above-concluded relative orders of
stability for both I-IV and V-VIII are in line with the maximal $\pi -$
energy of 1,1-divinyl isomers of polyenes among the branched ones
established in Ref. [35].

It is seen, therefore, that fourth order energies are sufficient to
distinguish between relative stabilities of representatives of different
principal series of polyenes. In the case of distinct isomers characterized
by more similar overall constitutions, however, the fourth order energies
often are uniform and, consequently, sixth order corrections should be
invoked. Let turn now to relevant examples.

Let us start with comparison of the above-discussed isomers of decapentaene
VI and IX containing the same number of CP(3)s equal to three. As it may be
easily proven after constructing the relevant principal matrices, a zero
fourth order energy is peculiar to both isomers under comparison and this
result evidently causes little surprise. Again, distinct numbers of CP(4)s
(namely 2 and 1 for systems VI and IX, respectively) allow us to expect the
sixth order energies to be responsible for different stabilities of these
isomers (Subsect. 3.2). Thus, let us now turn to corrections $\mathcal{E}%
_{(6)}(VI)$ and $\mathcal{E}_{(6)}(IX)$.

Matrices $\mathbf{G}_{(3)}^{o}(VI)$\ and $\mathbf{G}_{(3)}^{o}(IX)$\ along
with the consequent energy increments $\mathcal{E}_{(6)1}^{(+)}(VI)$\ and $%
\mathcal{E}_{(6)1}^{(+)}(IX)$\ are shown in Eq.(23) and discussed nearby. A
higher value of $\mathcal{E}_{(6)1}^{(+)}(VI)$\ vs. $\mathcal{E}%
_{(6)1}^{(+)}(IX)$ was in line with the relevant numbers of CP(4)s. Further,
matrices $\mathbf{G}_{(1)}\mathbf{G}_{(1)}^{+}\mathbf{G}_{(1)}(VI)$\ and $%
\mathbf{G}_{(1)}\mathbf{G}_{(1)}^{+}\mathbf{G}_{(1)}(IX)$\ determining the
second stabilizing increments $\mathcal{E}_{(6)2}^{(+)}(VI)$ and $\mathcal{E}%
_{(6)2}^{(+)}(IX),$\ respectively, prove to be similar in respect of
absolute values of their non-zero elements. This implies the numbers of the
relevant composite conjugated paths (consisting of three CP(2)s) to be
uniform in the isomers VI and IX. As a result, energy increments $\mathcal{E}%
_{(6)2}^{(+)}(VI)$ and $\mathcal{E}_{(6)2}^{(+)}(IX)$\ also coincide one
with another and equal to $40\gamma ^{6}/256.$

As with the above-discussed matrices $\mathbf{G}_{(3)}^{o}(VI)$\ and $%
\mathbf{G}_{(3)}^{o}(IX)$\ of Eq.(23), the matrix $\mathbf{G}_{(1)}\mathbf{G}%
_{(2)}^{+}(VI)$\ also contains more non-zero elements as compared to $%
\mathbf{G}_{(1)}\mathbf{G}_{(2)}^{+}(IX)$\ [Apart from a single element
equal to 2, these matrices involve eight and seven elements, respectively,
that are equal to either $1$ or $-$1. This result is in line with the
above-foreseen parallelism between alterations in elements of matrices $%
\mathbf{G}_{(3)}^{o}$ and $\mathbf{G}_{(1)}\mathbf{G}_{(2)}^{+}$\ (see the
last paragraph of Sect. 3)]. Consequently, the absolute value of the sixth
order destabilizing increment $\mathcal{E}_{(6)}^{(-)}(VI)$\ also exceeds
that of $\mathcal{E}_{(6)}^{(-)}(IX)$. Moreover, the same refers also to
increments $\mathcal{E}_{(6)}^{(u)}(VI)$\ and $\mathcal{E}_{(6)}^{(u)}(IX),$%
\ viz.%
\begin{equation}
\mathcal{E}_{(6)}^{(-)}(VI)\ =-\frac{96\gamma ^{6}}{256},\quad \mathcal{E}%
_{(6)}^{(-)}(IX)=-\frac{88\gamma ^{6}}{256},\quad \mathcal{E}%
_{(6)}^{(u)}(VI)\ =\frac{16\gamma ^{6}}{256},\quad \mathcal{E}%
_{(6)}^{(u)}(IX)=\frac{8\gamma ^{6}}{256}.
\end{equation}%
As is seen after summing up the relevant contributions, the sixth order
stabilization energy increases by $18\gamma ^{6}/256,$ whereas the absolute
value of destabilization grows only by $8\gamma ^{6}/256$\ when passing from
IX to VI. The total sixth order energies then coincide with zero and $%
-10\gamma ^{6}/256$ for isomers VI and IX, respectively, and indicate the
former pi-electron system to be more stable than the latter in accordance
with graph-theoretical conclusions of Ref.[35]. The present result may be
traced back to the relevant numbers of CP(4)s. The relation between these
numbers and relative stabilities of isomers VI and IX, however, is far from
being of a straightforward nature as the above discussion shows.

Let us return again to the semi-conjugated isomer of decapentaene (VII) and
compare it to a similar one (X) (Fig. 2). As with the above-considered
couple (VI and IX), the isomers VII and X also are characterized by the same
numbers of CP(3)s and, consequently, by uniform fourth order energies equal
to $-4\gamma ^{4}/64$. Different numbers of CP(4)s of these hydrocarbons
also deserve mention here (these coincide with 1 and 0 for isomers VII and
X, respectively). Thus, let us turn to the sixth order energies $\mathcal{E}%
_{(6)}(VII)$\ and $\mathcal{E}_{(6)}(X).$\ 

Matrices $\mathbf{G}_{(3)}^{o}(VII)$\ and $\mathbf{G}_{(3)}^{o}(X)$ exhibit
a clear parallelism between absolute values of their non-zero elements and
the respective numbers of CP(4)s as previously. Accordingly, the total
number of non-zero elements is higher in the matrix $\mathbf{G}_{(1)}\mathbf{%
G}_{(2)}^{+}(VII)$ as compared to\ $\mathbf{G}_{(1)}\mathbf{G}_{(2)}^{+}(X)$%
. Meanwhile, matrices $\mathbf{G}_{(1)}\mathbf{G}_{(1)}^{+}\mathbf{G}%
_{(1)}(VII)$\ and $\mathbf{G}_{(1)}\mathbf{G}_{(1)}^{+}\mathbf{G}_{(1)}(X)$
contain analogous non-zero elements. The total sixth order energies $%
\mathcal{E}_{(6)}(VII)$\ and $\mathcal{E}_{(6)}(X)$ then correspondingly
equal to $2\gamma ^{6}/256$ and $-8\gamma ^{6}/256.$ Thus, the isomer VII is
predicted to be more stable as compared to X in analogy with the
above-considered couple VI and IX. The decisive role of CP(4)s in the
formation of this result also is beyond any doubt.

Let us now compare the semi-conjugated system VIII to a closely related one
(XI). These isomers also are characterized by coinciding numbers of CP(3)s
and by uniform fourth order energies (equal to $-12\gamma ^{4}/64$). As
opposed to previous examples, however, both VIII and XI contain no CP(4)s.
This implies that the standard model of conjugated paths is not able to
distinguish between their stabilities. Again, the same isomers VIII and XI
differ one from another in numbers of semi-conjugated paths (SCP(4)s)
defined in Sect. 3. Indeed, the system VIII contains a single SCP(4)
embracing the C=C bonds under numbers 2,3,4,5. Meanwhile, the remaining
isomer XI involves two SCP(4)s made up of C=C bonds 1,2,3,4 and 2,3,4,5. In
this connection, comparison of sixth order energies of polyenes VIII and IX
is of particular interest.

As with the above-considered couples of isomers, matrices $\mathbf{G}_{(1)}%
\mathbf{G}_{(1)}^{+}\mathbf{G}_{(1)}(VIII)$\ and $\mathbf{G}_{(1)}\mathbf{G}%
_{(1)}^{+}\mathbf{G}_{(1)}(XI)$\ contain uniform sets of non-zero elements
and contribute coinciding increments to the relevant sixth order energies.
Again, matrices $\mathbf{G}_{(3)}^{o}(VIII)$\ and $\mathbf{G}_{(3)}^{o}(XI)$%
\ correspondingly involve six and eight non-zero elements equal to either $1$
or $-1$\ and thereby reflect different numbers of SCP(4)s present in these
systems. Consequently, the energy increments $\mathcal{E}_{(6)1}^{(+)}(VIII)%
\ $and\ $\mathcal{E}_{(6)1}^{(+)}(XI)$\ coincide with $6\gamma ^{6}/256$\
and $8\gamma ^{6}/256,$\ respectively. Thus, $\mathcal{E}_{(6)1}^{(+)}(XI)$\
exceeds $\mathcal{E}_{(6)1}^{(+)}(VIII)$\ by $2\gamma ^{6}/256$\ in
accordance with the expectation. Analogously, matrices $\mathbf{G}_{(1)}%
\mathbf{G}_{(2)}^{+}(VIII)$\ and $\mathbf{G}_{(1)}\mathbf{G}_{(2)}^{+}(XI)$\
contain three and four unit elements, respectively, and yield the following
energy increments \ 
\begin{equation}
\mathcal{E}_{(6)}^{(-)}(VIII)\ =-\frac{24\gamma ^{6}}{256},\quad \mathcal{E}%
_{(6)}^{(-)}(XI)=-\frac{32\gamma ^{6}}{256},\quad \mathcal{E}%
_{(6)}^{(u)}(VIII)=\mathcal{E}_{(6)}^{(u)}(XI)=0.
\end{equation}%
Thus, the absolute value of the destabilizing increment is increased by $%
8\gamma ^{6}/256$\ when passing from VIII to XI and this alteration exceeds
that of the stabilizing increment considerably. It is no surprise in this
connection that the isomer VIII proves to be more stable as compared to its
counterpart XI [The total sixth order energies correspondingly equal to $%
10\gamma ^{6}/256$ and $4\gamma ^{6}/256$]. Such a somewhat unexpected
result may be entirely traced back to the higher destabilizing effect of
composite conjugated paths (CCP(4)s) underlying matrices $\mathbf{G}_{(1)}%
\mathbf{G}_{(2)}^{+}$\ in the isomer XI vs. VIII.

\begin{figure}
\includegraphics[width=0.6\textwidth]{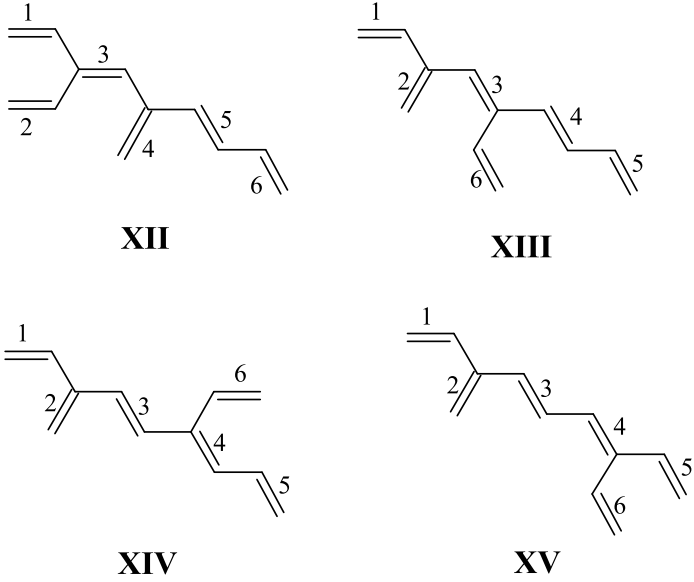}
\caption{Selected polyenes (XII-XV) containing six C=C bonds (N=6).
Numberings of these bonds also are shown}
\end{figure}

The last example under our interest embraces four isomers XII, XIII, XIV and
XV of Fig. 3, all of them containing six C=C bonds (N=6) and characterized
by uniform fourth order energies $-6\gamma ^{4}/64$ in addition. The total
numbers of CP(3)s also are uniform here and coincide with 3. Meanwhile, the
relevant numbers of the standard CP(4)s correspondingly equal to 0, 1, 1,
and 2. This implies the isomers XIII and XIV to be of the same composition
in terms of the standard conjugated paths. [It is no surprise that these
isomers have been never discriminated as concluded in Ref.[5]]. The relevant
numbers of SCP(4)s coincide with 3,2,1 and 1, respectively.

Separate increments to the sixth order energies of the above-enumerated
systems also follow the above-observed trends. Thus, the first stabilizing
increments ($\mathcal{E}_{(6)1}^{(+)}$) originating from matrices
$\mathbf{G}_{(3)}^{o}$\ are as follows
\begin{equation}
\mathcal{E}_{(6)1}^{(+)}(XII)\ =\frac{22\gamma ^{6}}{256},\mathcal{E}%
_{(6)1}^{(+)}(XIII)=\frac{34\gamma ^{6}}{256},\mathcal{E}_{(6)1}^{(+)}(XIV)\
=\frac{32\gamma ^{6}}{256},\mathcal{E}_{(6)1}^{(+)}(XV)=\frac{42\gamma ^{6}}{%
256}
\end{equation}%
and correlate with the total numbers of both CP(4)s and SCP(4)s.
Furthermore, matrices $\mathbf{G}_{(1)}\mathbf{G}_{(1)}^{+}\mathbf{G}_{(1)}$
are similar as previously except for the matrix $\mathbf{G}_{(1)}\mathbf{G}%
_{(1)}^{+}\mathbf{G}_{(1)}(XIII)$\ possessing a somewhat higher number of
non-zero elements as compared to the remaining ones. The relevant energy
increments the take the form\ 
\begin{equation}
\mathcal{E}_{(6)2}^{(+)}(XII)\ =\mathcal{E}_{(6)2}^{(+)}(XIV)=\mathcal{E}%
_{(6)2}^{(+)}(XV)=\frac{50\gamma ^{6}}{256},\quad \mathcal{E}%
_{(6)2}^{(+)}(XIII)=\frac{53\gamma ^{6}}{256}.
\end{equation}%
\ Finally, matrices $\mathbf{G}_{(1)}\mathbf{G}_{(2)}^{+}$\ contain
different numbers of unit elements along with a single element 2, namely
seven, nine, eight and nine unit elements for isomers XII-XIV, respectively.
These matrices give birth to following energy increments \ 
\begin{eqnarray}
\mathcal{E}_{(6)}^{(-)}(XII) &=&-\frac{88\gamma ^{6}}{256},\;\mathcal{E}%
_{(6)}^{(-)}(XIII)=\mathcal{E}_{(6)}^{(-)}(XV)=-\frac{104\gamma ^{6}}{256},\;%
\mathcal{E}_{(6)}^{(-)}(XIV)=-\frac{96\gamma ^{6}}{256},  \notag \\
\mathcal{E}_{(6)}^{(u)}(XII) &=&0,\quad \mathcal{E}_{(6)}^{(u)}(XIII)=%
\mathcal{E}_{(6)}^{(u)}(XIV)=\frac{8\gamma ^{6}}{256},\quad \mathcal{E}%
_{(6)}^{(u)}(XV)=\frac{16\gamma ^{6}}{256}.
\end{eqnarray}%
It is seen that absolute values of destabilizing increments also correlate
with total numbers of both CP(4)s and SCP(4)s in the isomers XII-XIV. After
summing up all increments concerned, we obtain 
\begin{equation}
\mathcal{E}_{(6)}(XII)\ =-\frac{16\gamma ^{6}}{256},\;\mathcal{E}%
_{(6)}(XIII)=-\frac{9\gamma ^{6}}{256},\;\mathcal{E}_{(6)}(XIV)\ =-\frac{%
6\gamma ^{6}}{256},\;\mathcal{E}_{(6)}(XV)=\frac{4\gamma ^{6}}{256}
\end{equation}%
Thus, the relative stability grows with the increasing number of the
standard CP(4)s in this case too. Moreover, the isomers XIII and XIV (both
containing a single CP(4)) also are discriminated when applying the
perturbative approach: The isomer XIII is predicted to be less stable as
compared to XIV as it was the case with XI vs. VIII.

\vspace{4mm}

\section{Conclusions}

Analysis of power series for total energies of pi-electron systems of
acyclic conjugated hydrocarbons (polyenes) supports the principal
assumptions underlying the model of conjugated paths and thereby offers a
justification of the latter. In this respect, the following points may be
mentioned:

(i) The standard conjugated paths (CPs) embracing two, three and four
linearly-connected C=C bonds (CP(2)s, CP(3)s and CP(4)s) contribute
significantly to terms of power series for total energies of the second,
fourth and sixth orders, respectively;

(ii) The afore-mentioned contributions always are of positive signs (in
negative energy units) and thereby of stabilizing nature;

(iii) Relative values of the contributions concerned depend upon total
numbers of the respective standard CPs present in the given hydrocarbon. In
particular, contributions of CP(2)s and CP(3)s to energy corrections of the
second and fourth orders, respectively, are expressible as sums of
transferable increments of individual CPs and thereby these are directly
proportional to the numbers of the latter; 

(iv) The decisive energy correction of the second order (coinciding with the
Dewar energy of the PMO theory [14]) is determined exclusively by the number
of the simplest conjugated paths (CP(2)s) embracing two C=C bonds connected
by a C-C bond.

Again, application of the perturbative approach to relative stabilities of
pi-electron systems of polyenes undertaken in the above study contributes to
an extension of the very concept and/or model of conjugated paths. This
conclusion is based on the following properties of the power series for
total energies:

(i) Members of the series of the fourth ($\mathcal{E}_{(4)}$) and sixth
orders ($\mathcal{E}_{(6)}$) contain both positive (stabilizing) and
negative (destabilizing) components and these are in some relation one with
another in addition. This implies that destabilizing factors also manifest
themselves in polyenes that are able to play an equally decisive role in the
formation of the final total energy;

(ii) Negative (destabilizing) components of the energy corrections $\mathcal{%
E}_{(4)}$ and $\mathcal{E}_{(6)}$\ are interpretable as contributions of
conjugated paths of a non-standard (composite) nature defined as successive
products of two or three connected standard CP(2)s and/or CP(3)s. Moreover,
the self-returning composite conjugated paths (SRCCPs) prove to be
especially important, wherein an even number of C=C bonds is involved and
each of them is visited twice;

(iii) The usual linear conjugated paths embracing four C=C bonds (CP(4)s)
are not the only fragments (substructures) participating in the formation of
positive (stabilizing) components of the sixth order energy. In particular,
four-membered fragments containing both a linearly-conjugated segment and a
cross-conjugated one [the so-called semi-conjugated paths (SCP(4)s] also
contribute to the sixth order stabilization of polyenes; 

(iv) The energy correction of the sixth order generally is a non-additive
quantity with respect to increments of individual participating fragments
(substructures) including the standard CPs. Consequently, the total value of
this correction depends upon the actual mutual arrangement of these
fragments in the given hydrocarbon.

Consideration of specific examples also corroborates the extended nature
and, consequently, a higher discriminative potential of the perturbative
approach applied vs. the usual CP model. In this respect, the most important
conclusions are as follows:

(i) Destabilizing increments of self-returning composite conjugated paths
(SRCCPs) prove to be generally responsible for lower relative stabilities of
branched isomers of polyenes vs. their linear counterparts; 

(ii) isomers of extended polyenes containing different numbers of
semi-conjugated paths (SCPs) usually are represented by distinct sixth order
energies even if the relevent numbers of the standard CPs are uniform.

\appendix

\begin{center}
\Large\textbf{Appendix}
\end{center}

\section{Derivation of expressions for the sixth order energy corrections}

In its most general form, the power series for total energies of molecules
and molecular systems has been originally derived in Ref.[17]. This study
contains members of the power series up to fourth order ($k=4$). The
relevant fifth order terms may be found in Ref.[15]. A direct extension of
the above-cited derivation to terms of higher orders (including the sixth
order ones) is a rather cumbersome procedure. In this connection, we will
confine ourselves here to a less general Hamiltonian matrix vs. that of
Refs. [17,18] as described below. Nevertheless, the overall methodology to
be invoked closely resembles the original one [17]. The main points of the
latter are as follows: First, the interrelation [36] is employed between the
total energy being sought ($\mathcal{E}$), the Hamiltonian matrix of the
system(s) concerned ($\mathbf{H}$) and the relevant representation of the
one-electron density matrix (the charge- bond order (CBO) matrix) $\mathbf{P,%
}$ viz. 
\begin{equation}
\mathcal{E}=Tr(\mathbf{PH}).  \tag{A1}
\end{equation}%
Second, the matrix $\mathbf{P}$\ is derived directly [18] on the basis of
solution of the so-called commutation equation [36]. For Hamiltonian
matrices ($\mathbf{H}$) consisting of zero and first order members ($\mathbf{%
H}_{(0)}$ and $\mathbf{H}_{(1)}$, respectively), the above-mentioned
solution may be carried out perturbatively. As a result, both the CBO matrix 
$\mathbf{P}$\ and the total energy $\mathcal{E}$\ are expressible as sums of
corrections $\mathbf{P}_{(k)}$\ and $\mathcal{E}_{(k)}$\ of increasing
orders ($k$). Moreover, each energy correction $\mathcal{E}_{(k)}$\ is
additionally representable as a sum of two components, viz.%
\begin{equation}
\mathcal{E}_{(k)}=\mathcal{E}_{(k)}^{(\alpha )}+\mathcal{E}_{(k)}^{(\beta
)},\quad \mathcal{E}_{(k)}^{(\alpha )}=Tr(\mathbf{P}_{(k)}\mathbf{H}%
_{(0)}),\quad \mathcal{E}_{(k)}^{(\beta )}=Tr(\mathbf{P}_{(k-1)}\mathbf{H}%
_{(1)}).  \tag{A2}
\end{equation}

The most general Hamiltonian matrix ($\mathbf{H}$) underlying the original
derivation of $\mathcal{E}_{(k)}$\ [17] is as follows 
\begin{equation}
\mathbf{H}=\mathbf{H}_{(0)}+\mathbf{H}_{(1)}=\left\vert 
\begin{array}{cc}
\mathbf{E}_{(+)} & \mathbf{0} \\ 
\mathbf{0} & \mathbf{-E}_{(-)}%
\end{array}%
\right\vert +\left\vert 
\begin{array}{cc}
\mathbf{S} & \mathbf{R} \\ 
\mathbf{R}^{+} & \mathbf{Q}%
\end{array}%
\right\vert ,  \tag{A3}
\end{equation}%
where $\mathbf{E}_{(+)},\mathbf{E}_{(-)},\mathbf{S,R}$ and $\mathbf{Q}$ are
certain $N\times N-$dimensional submatrices. Systems underlying the matrix $%
\mathbf{H}$\ and details of its construction (see e.g.[15,17]) are of no
importance here. Let us note only that the relevant $2N-$dimensional basis
set $\{\Psi \}$\ is assumed to consist of two well-separated $N-$dimensional
subsets $\{\Psi _{(+)}\}$\ and $\{\Psi _{(-)}\}.$\ The minus sign in front
of $\mathbf{E}_{(-)}$ of Eq.(A3) is introduced for convenience. The
superscript + designates the transposed (Hermitian- conjugate) matrix.

The above-exhibited form of the initial Hamiltonian matrix $\mathbf{H}$\
allowed us to look for the CBO matrix $\mathbf{P}$, separate members of the
power series of which ($\mathbf{P}_{(k)}$) also are divisible into four
submatrices (blocks). Moreover, a new version of the Rayleigh-Schr\"{o}%
dinger perturbation theory (RSPT)\ has been formulated, wherein entire
submatrices (blocks) of the matrix $\mathbf{H}$\ (i.e. non-commutative
quantities) play the central role instead of usual (commutative) matrix
elements. Accordingly, the new PT [19,20] has been called non-commutative
RSPT (NCRSPT).\ As a result of its application, the corrections $\mathbf{P}%
_{(k)}$\ take the following form 
\begin{equation}
\mathbf{P}_{(k)}=-2\left\vert 
\begin{array}{cc}
\mathbf{X}_{(k)+} & \mathbf{G}_{(k)} \\ 
\mathbf{G}_{(k)}^{+} & \mathbf{-X}_{(k)-}%
\end{array}%
\right\vert ,  \tag{A4}
\end{equation}%
where $\mathbf{G}_{(k)}$\ are the so-called principal matrices of the NCRSPT
determined by certain matrix equations [18-20]. Meanwhile, diagonal
positions of the corrections $\mathbf{P}_{(k)}$\ are occupied by matrices $%
\mathbf{X}_{(k)+}$\ and $\mathbf{X}_{(k)-}$\ that have been referred to as
intrasubset population matrices. These correspondingly refer to subsets $%
\{\Psi _{(+)}\}$\ and $\{\Psi _{(-)}\}$\ and are expressible in the form of
sums of products of matrices $\mathbf{G}_{(k)}$\ of lower orders as
exemplified below by Eqs. (A9) and (A15). The original derivation of members
($\mathbf{P}_{(k)}$) of power series for the matrix $\mathbf{P}$\ [18]
embraced terms to within second order only. Nevertheless, it is easily
extendable to any $k$.

As already mentioned, we confine ourselves here to a particular case of the
matrix $\mathbf{H}$\ of Eq.(A3). To this end, let us accept the equality $%
\mathbf{E}_{(+)}\mathbf{=E}_{(-)}=\mathbf{I}$. [Besides, the resulting
simplified Hamiltonian matrix coincides with that following from the initial
matrix of our study shown in Eq.(1) after transforming the latter into the
basis of bond orbitals (BOs) $\{\varphi \}$[15]. (submatrices $\mathbf{I}$
and $\mathbf{-I}$ correspondingly represent one-electron energies of bonding
BOs (BBOs) and of antibonding BOs (ABOs)]. The most important advantage of
the above-specified condition consists in the possibility of an algebraic
solution of matrix equations determining the principal matrices of the
NCRSPT $\mathbf{G}_{(k)},k=1,2,3...$. As a result, these matrices meet the
following recurrence relations%
\begin{equation}
\mathbf{G}_{(k)}=-\frac{1}{2}(\mathbf{SG}_{(k-1)}-\mathbf{G}_{(k-1)}\mathbf{%
Q)-L}_{(k)},  \tag{A5}
\end{equation}%
where $\mathbf{L}_{(k)}$\ are products of the same matrices of lower orders,
e. g. 
\begin{align}
\mathbf{L}_{(1)} =&\mathbf{L}_{(2)}=\mathbf{0,\quad L}_{(3)}=2\mathbf{G}
_{(1)}\mathbf{G}_{(1)}^{+}\mathbf{G}_{(1)},  \notag \\
\mathbf{L}_{(4)} =&\mathbf{G}_{(2)}\mathbf{G}_{(1)}^{+}\mathbf{G}_{(1)}+2
\mathbf{G}_{(1)}\mathbf{G}_{(2)}^{+}\mathbf{G}_{(1)}+\mathbf{G}_{(1)}
\mathbf{G}_{(1)}^{+}\mathbf{G}_{(2)}\,,\mathrm{etc.}  \tag{A6}
\end{align}
Finally, a useful relation $\mathbf{R}=\mathbf{-}2\mathbf{G}_{(1)}$\ follows
for first order matrices in this case. Substituting the latter relation
along with Eqs.(A3) and (A4) into Eq.(A2) yields the following expressions
for separate components of the sixth order energy, viz.%
\begin{align}
\mathcal{E}_{(6)}^{(\alpha )} =&-4Tr(\mathbf{X}_{(6)+}),  \tag{A7} \\
\quad \mathcal{E}_{(6)}^{(\beta )} =&-2Tr(\mathbf{X}_{(5)+}\mathbf{S-X}
_{(5)-}\mathbf{Q)+}8Tr(\mathbf{G}_{(5)}\mathbf{G}_{(1)}^{+}).  \tag{A8}
\end{align}
Let us now consider the components of Eqs.(A7) and (A8) separately. The
matrix $\mathbf{X}_{(6)+}$\ determining the first component $\mathcal{E}%
_{(k)}^{(\alpha )}$\ is expressible as follows%
\begin{align}
\mathbf{X}_{(6)+} =&\mathbf{G}_{(5)}\mathbf{G}_{(1)}^{+}+\mathbf{G}_{(1)}
\mathbf{G}_{(5)}^{+}+\mathbf{G}_{(4)}\mathbf{G}_{(2)}^{+}+\mathbf{G}_{(2)}
\mathbf{G}_{(4)}^{+}+\mathbf{G}_{(3)}\mathbf{G}_{(3)}^{+}+  \notag \\
&+2\mathbf{G}_{(1)}\mathbf{G}_{(1)}^{+}\mathbf{G}_{(1)}\mathbf{G}_{(1)}^{+}
\mathbf{G}_{(1)}\mathbf{G}_{(1)}^{+}+\mathbf{G}_{(1)}\mathbf{G}_{(1)}^{+}
\mathbf{G}_{(1)}\mathbf{G}_{(3)}^{+}+  \notag \\
&+\mathbf{G}_{(1)}\mathbf{G}_{(1)}^{+}\mathbf{G}_{(3)}\mathbf{G}_{(1)}^{+}+
\mathbf{G}_{(1)}\mathbf{G}_{(3)}^{+}\mathbf{G}_{(1)}\mathbf{G}_{(1)}^{+}+
\mathbf{G}_{(3)}\mathbf{G}_{(1)}^{+}\mathbf{G}_{(1)}\mathbf{G}_{(1)}^{+}+ 
\notag \\
&+\mathbf{G}_{(1)}\mathbf{G}_{(1)}^{+}\mathbf{G}_{(2)}\mathbf{G}_{(2)}^{+}+
\mathbf{G}_{(2)}\mathbf{G}_{(2)}^{+}\mathbf{G}_{(1)}\mathbf{G}_{(1)}^{+}+
\mathbf{G}_{(1)}\mathbf{G}_{(2)}^{+}\mathbf{G}_{(1)}\mathbf{G}_{(2)}^{+}+ 
\notag \\
&+\mathbf{G}_{(2)}\mathbf{G}_{(1)}^{+}\mathbf{G}_{(1)}\mathbf{G}_{(2)}^{+}+
\mathbf{G}_{(1)}\mathbf{G}_{(2)}^{+}\mathbf{G}_{(2)}\mathbf{G}_{(1)}^{+}+
\mathbf{G}_{(2)}\mathbf{G}_{(1)}^{+}\mathbf{G}_{(2)}\mathbf{G}_{(1)}^{+}. 
\tag{A9}
\end{align}
Substituting the above formula into Eq.(A7) shows that the component
$\mathcal{E}_{(6)}^{(\alpha )}$ contains matrices $\mathbf{G}_{(k)}$\ up to
$k=5$. The next step then consists in eliminating the matrices
$\mathbf{G}_{(5)}$ and $\mathbf{G}_{(4)}$ from the expression concerned on the basis
of our previous experience when dealing with similar relations. To this end,
let us take $Tr(\mathbf{G}_{(5)}\mathbf{G}_{(1)}^{+})$ and
$Tr(\mathbf{G}_{(4)}\mathbf{G}_{(2)}^{+})$ separately and substitute recurrence
relations of Eq.(A5) for $\mathbf{G}_{(5)}$ and $\mathbf{G}_{(4)}$,
respectively. Moreover, Eq.(A6) also should be used along with allowed
cyclic transpositions of matrices inside the $Trace$ signs. The results of
these procedures are as follows
\begin{align}
Tr(\mathbf{G}_{(5)}\mathbf{G}_{(1)}^{+}) =&Tr(\mathbf{G}_{(3)}\mathbf{G}
_{(3)}^{+})-2Tr(\mathbf{G}_{(3)}\mathbf{G}_{(1)}^{+}\mathbf{G}_{(1)}
\mathbf{G}_{(1)}^{+})-2Tr(\mathbf{G}_{(1)}\mathbf{G}_{(1)}^{+}\mathbf{G}_{(2)}
\mathbf{G}_{(2)}^{+})  \notag \\
&-2Tr(\mathbf{G}_{(1)}^{+}\mathbf{G}_{(1)}\mathbf{G}_{(2)}^{+}\mathbf{G}_{(2)})
-2Tr(\mathbf{G}_{(1)}\mathbf{G}_{(2)}^{+}\mathbf{G}_{(1)}
\mathbf{G}_{(2)}^{+})  \notag \\
&-2Tr(\mathbf{G}_{(1)}\mathbf{G}_{(1)}^{+}\mathbf{G}_{(1)}\mathbf{G}_{(1)}^{+}
\mathbf{G}_{(1)}\mathbf{G}_{(1)}^{+}),  \tag{A10}
\end{align}
\begin{align}
Tr(\mathbf{G}_{(4)}\mathbf{G}_{(2)}^{+}) =&Tr(\mathbf{G}_{(3)}\mathbf{G}_{(3)}^{+})
+2Tr(\mathbf{G}_{(3)}\mathbf{G}_{(1)}^{+}\mathbf{G}_{(1)}\mathbf{G}_{(1)}^{+})
-Tr(\mathbf{G}_{(1)}\mathbf{G}_{(1)}^{+}\mathbf{G}_{(2)}
\mathbf{G}_{(2)}^{+})  \notag \\
&-Tr(\mathbf{G}_{(1)}^{+}\mathbf{G}_{(1)}\mathbf{G}_{(2)}^{+}\mathbf{G}_{(2)})
-2Tr(\mathbf{G}_{(1)}\mathbf{G}_{(2)}^{+}\mathbf{G}_{(1)}
\mathbf{G}_{(2)}^{+})  \tag{A11}
\end{align}
Employment of Eqs.(A7) and (A9)-(A11) then yields the following formula
for $\mathcal{E}_{(6)}^{(\alpha )}$ 
\begin{align}
\mathcal{E}_{(6)}^{(\alpha )} =&-20Tr(\mathbf{G}_{(3)}\mathbf{G}_{(3)}^{+})
-16Tr(\mathbf{G}_{(3)}\mathbf{G}_{(1)}^{+}\mathbf{G}_{(1)}\mathbf{G}_{(1)}^{+})
+16Tr(\mathbf{G}_{(1)}\mathbf{G}_{(1)}^{+}\mathbf{G}_{(2)}
\mathbf{G}_{(2)}^{+})+  \notag \\
&+16Tr(\mathbf{G}_{(1)}^{+}\mathbf{G}_{(1)}\mathbf{G}_{(2)}^{+}\mathbf{G}_{(2)})
+24Tr(\mathbf{G}_{(1)}\mathbf{G}_{(2)}^{+}\mathbf{G}_{(1)}
\mathbf{G}_{(2)}^{+})  \notag \\
&+8Tr(\mathbf{G}_{(1)}\mathbf{G}_{(1)}^{+}\mathbf{G}_{(1)}\mathbf{G}_{(1)}^{+}
\mathbf{G}_{(1)}\mathbf{G}_{(1)}^{+})  \tag{A12}
\end{align}
that, in turn, may be simplified considerably after eliminating
$Tr(\mathbf{G}_{(3)}\mathbf{G}_{(1)}^{+}\mathbf{G}_{(1)}\mathbf{G}_{(1)}^{+})$.
To this end, the following relation should be used, viz.
\begin{align}
Tr(\mathbf{G}_{(3)}\mathbf{G}_{(1)}^{+}\mathbf{G}_{(1)}\mathbf{G}_{(1)}^{+})
=&Tr(\mathbf{G}_{(1)}\mathbf{G}_{(1)}^{+}\mathbf{G}_{(2)}\mathbf{G}_{(2)}^{+})
-2Tr(\mathbf{G}_{(1)}\mathbf{G}_{(1)}^{+}\mathbf{G}_{(1)}\mathbf{G}_{(1)}^{+}
\mathbf{G}_{(1)}\mathbf{G}_{(1)}^{+})  \notag \\
&+Tr(\mathbf{G}_{(1)}^{+}\mathbf{G}_{(1)}\mathbf{G}_{(2)}^{+}\mathbf{G}_{(2)})
-Tr(\mathbf{G}_{(1)}\mathbf{G}_{(2)}^{+}\mathbf{G}_{(1)}
\mathbf{G}_{(2)}^{+}).  \tag{A13}
\end{align}
Derivation of Eq.(A13) may be carried out analogously to those of Eqs.(A10)
and (A11). The final expression for $\mathcal{E}_{(6)}^{(\alpha )}$ is then
as follows
\begin{equation}
\mathcal{E}_{(6)}^{(\alpha )}=-20Tr(\mathbf{G}_{(3)}\mathbf{G}_{(3)}^{+})
+40Tr(\mathbf{G}_{(1)}\mathbf{G}_{(2)}^{+}\mathbf{G}_{(1)}\mathbf{G}_{(2)}^{+})
+40Tr(\mathbf{G}_{(1)}\mathbf{G}_{(1)}^{+}\mathbf{G}_{(1)}
\mathbf{G}_{(1)}^{+}\mathbf{G}_{(1)}\mathbf{G}_{(1)}^{+}).  \tag{A14}
\end{equation}

The second component of the sixth order energy ($\mathcal{E}_{(6)}^{(\beta )}
$) also may be reformulated similarly. The first step of the relevant
procedure consists in substituting into Eq.(A8) the expressions for
$\mathbf{X}_{(5)+}$ and $\mathbf{X}_{(5)-}$, viz. 
\begin{align}
\mathbf{X}_{(5)+} =&\mathbf{G}_{(4)}\mathbf{G}_{(1)}^{+}+\mathbf{G}_{(1)}
\mathbf{G}_{(4)}^{+}+\mathbf{G}_{(3)}\mathbf{G}_{(2)}^{+}+\mathbf{G}_{(2)}
\mathbf{G}_{(3)}^{+}+\mathbf{G}_{(1)}\mathbf{G}_{(1)}^{+}\mathbf{G}_{(1)}
\mathbf{G}_{(2)}^{+}+  \notag \\
&\mathbf{G}_{(1)}\mathbf{G}_{(1)}^{+}\mathbf{G}_{(2)}\mathbf{G}_{(1)}^{+}+
\mathbf{G}_{(1)}\mathbf{G}_{(2)}^{+}\mathbf{G}_{(1)}\mathbf{G}_{(1)}^{+}+
\mathbf{G}_{(2)}\mathbf{G}_{(1)}^{+}\mathbf{G}_{(1)}\mathbf{G}_{(1)}^{+}, 
\tag{A15}
\end{align}
\begin{align}
\mathbf{X}_{(5)-} =&\mathbf{G}_{(4)}^{+}\mathbf{G}_{(1)}+\mathbf{G}_{(1)}^{+}
\mathbf{G}_{(4)}+\mathbf{G}_{(3)}^{+}\mathbf{G}_{(2)}+\mathbf{G}_{(2)}^{+}
\mathbf{G}_{(3)}+\mathbf{G}_{(1)}^{+}\mathbf{G}_{(1)}\mathbf{G}_{(1)}^{+}
\mathbf{G}_{(2)}+  \notag \\
&\mathbf{G}_{(1)}^{+}\mathbf{G}_{(1)}\mathbf{G}_{(2)}^{+}
\mathbf{G}_{(1)}++\mathbf{G}_{(1)}^{+}\mathbf{G}_{(2)}\mathbf{G}_{(1)}^{+}
\mathbf{G}_{(1)}+\mathbf{G}_{(2)}^{+}\mathbf{G}_{(1)}\mathbf{G}_{(1)}^{+}
\mathbf{G}_{(1)}. 
\tag{A16}
\end{align}
Thereupon, relations of Eqs.(A10), (A11) and (A13) should be invoked to
eliminate $Tr(\mathbf{G}_{(5)}\mathbf{G}_{(1)}^{+})$,
$Tr(\mathbf{G}_{(4)}\mathbf{G}_{(2)}^{+})$ and
$Tr(\mathbf{G}_{(3)}\mathbf{G}_{(1)}^{+}
\mathbf{G}_{(1)}\mathbf{G}_{(1)}^{+})$, respectively. The result is as follows
\begin{equation}
\mathcal{E}_{(6)}^{(\beta )}=24Tr(\mathbf{G}_{(3)}\mathbf{G}_{(3)}^{+})-48Tr(
\mathbf{G}_{(1)}\mathbf{G}_{(2)}^{+}\mathbf{G}_{(1)}\mathbf{G}_{(2)}^{+})
-48Tr(\mathbf{G}_{(1)}\mathbf{G}_{(1)}^{+}\mathbf{G}_{(1)}\mathbf{G}_{(1)}^{+}
\mathbf{G}_{(1)}\mathbf{G}_{(1)}^{+}).  \tag{A17}
\end{equation}
After summing up the two components of the sixth order energy shown in
Eqs.(A14) and (A17) in accordance with Eq.(A2), we finally obtain 
\begin{equation}
\mathcal{E}_{(6)}=4Tr(\mathbf{G}_{(3)}\mathbf{G}_{(3)}^{+})-8Tr(\mathbf{G}_{(1)}
\mathbf{G}_{(2)}^{+}\mathbf{G}_{(1)}\mathbf{G}_{(2)}^{+})-8Tr(\mathbf{G}_{(1)}
\mathbf{G}_{(1)}^{+}\mathbf{G}_{(1)}\mathbf{G}_{(1)}^{+}\mathbf{G}_{(1)}
\mathbf{G}_{(1)}^{+}).  \tag{A18}
\end{equation}

The expression of Eq.(18) seems to be the most compact form of the
correction concerned. However, it is not the most convenient one for
practical applications, especially for the attempts of finding relations
between separate increments of the overall correction $\mathcal{E}_{(6)}$,
on the one hand, and individual interorbital interactions, on the other
hand. The main reason for that consists in the rather involved nature of the
matrix $\mathbf{G}_{(3)}$. Indeed, the relevant definition shown in
Eqs.(A5) and (A6) embraces the product $\mathbf{G}_{(1)}\mathbf{G}_{(1)}^{+}
\mathbf{G}_{(1)}$ (along with the $\mathbf{G}_{(2)}\mathbf{-}$containing
term) that gives birth to increments like that of the last term of Eq.(A18).
To be able to sum up these similar increments, let us define a new matrix
$\mathbf{G}_{(3)}^{o}$\ coinciding with the $\mathbf{G}_{(2)}-$containing
term of Eq.(A5) for $k=3$ as shown by the first relation of Eq.(12).
Accordingly, the matrix $\mathbf{G}_{(3)}$\ of Eq.(A18) may be replaced by
$\mathbf{G}_{(3)}^{o}-2\mathbf{G}_{(1)}\mathbf{G}_{(1)}^{+}\mathbf{G}_{(1)}$.
Thereupon, we may get rid of the newly-emerging term
$Tr(\mathbf{G}_{(3)}^{o}\mathbf{G}_{(1)}^{+}\mathbf{G}_{(1)}
\mathbf{G}_{(1)}^{+}\mathbf{)}$ by
constructing a relation like that of Eq.(A13). The final formula for
$\mathcal{E}_{(6)}$ is then as follows
\begin{align}
\mathcal{E}_{(6)} =&4Tr(\mathbf{G}_{(3)}^{o}\mathbf{G}_{(3)}^{o+})+8Tr(
\mathbf{G}_{(1)}\mathbf{G}_{(1)}^{+}\mathbf{G}_{(1)}\mathbf{G}_{(1)}^{+}
\mathbf{G}_{(1)}\mathbf{G}_{(1)}^{+})-16Tr(\mathbf{G}_{(1)}\mathbf{G}_{(1)}^{+}
\mathbf{G}_{(2)}\mathbf{G}_{(2)}^{+})  \notag \\
&-16Tr(\mathbf{G}_{(1)}^{+}\mathbf{G}_{(1)}\mathbf{G}_{(2)}^{+}\mathbf{G}_{(2)})
+8Tr(\mathbf{G}_{(1)}\mathbf{G}_{(2)}^{+}\mathbf{G}_{(1)}
\mathbf{G}_{(2)}^{+}).  \tag{A19}
\end{align}
Separate terms of the above expression are exhibited in Eqs.~(8)--(11) and
discussed nearby. It deserves adding here that third and fourth increments
of Eq.(A19) prove to be uniform in the case of AHs owing to the
skew-symmetric nature of matrices $\mathbf{G}_{(1)}$ and $\mathbf{G}_{(2)}$ [28].
After summing up these increments, a single destabilizing component
of the sixth order energy ($\mathcal{E}_{(6)}^{(-)}$) arises (see Eq.(10)).

\end{document}